\documentclass[amsmath,amssymb,aps,prfluids,]{revtex4-2}
\usepackage{graphicx}
\usepackage{epstopdf, epsfig}

\usepackage{amsmath}
\usepackage{comment}
\usepackage[export]{adjustbox}
\usepackage{xcolor}
\usepackage{hyperref}
\hypersetup{
    colorlinks = true,
    urlcolor   = blue,
    citecolor  = blue,
}
\newcommand{\overbar}[1]{\mkern 1.5mu\overline{\mkern-1.5mu#1\mkern-1.5mu}\mkern 1.5mu}

\begin{document}

\title{The characteristics of the meandering effect in a stratified wake}

\author{Xinyi Huang}
\affiliation{Graduate Aerospace Laboratories, California Institute of Technology, Pasadena, CA, 91125, USA}
\email{xinyih@caltech.edu}
\author{Jiaqi J. L. Li}
\affiliation{Department of Mechanical Engineering, The Pennsylvania State University, University Park, PA, 16802, USA}

\begin{abstract}

The gradual suppression of the vertical motions and the emergence of large-scale horizontal structures in the late wake are commonly seen in a stratified wake flow. 
We isolate the resulting wake meandering from the stationary velocity, i.e., the velocity without meandering, by utilizing direct numerical simulations covering bulk Reynolds numbers $Re_B=U_B D/\nu$ between $[10 000, 50 000]$ and bulk Froude number $Fr_B=U_B/ND$ between $[2, 50]$ ($U_B$ is the freestream velocity, $D$ is the characteristic length scale, and $N$ is the buoyancy frequency). 
The meandering range is growing in the horizontal direction as the wake width, but decreases in the vertical direction, different from the wake height. 
The meandering, especially in the horizontal direction, leads to the distortion of the instantaneous velocity and layered flow structures. 
Thus, the mean velocity profile deviates from the assumption of two-dimensional self-similarity in the late wake. 
Apart from the distortion of the velocity profile, the meandering of the wake also differentiates the statistics of the horizontal direction from those of the vertical direction. 
The scaling of the width and the height transitions at different points, and they are related to the scaling of the velocity deficit through the momentum flux equation. 
Through theoretical analysis, we obtain the superposition of the meandering to the stationary velocity. 
We can accurately measure how the meandering distorts the self-similar velocity profile and impacts the scaling of the width and height, and thus how the meandering changes the scaling of the velocity deficit. 
\end{abstract}

\maketitle

\section{Introduction}
\label{sec:intro}

Stratified flows are ubiquitous in nature, both in the atmosphere \cite{mahrt1999stratified} and in the ocean \cite{hopfinger1987turbulence,yin2023modeling}.
Their impacts are significant in various engineering and geophysical applications, such as the wake flow of aircraft, wind turbines, and underwater submersibles.
The existence of stratification leads to long-lived structures and allows for tracing its source \cite{spedding2014wake}. 
The wake behind a bluff or slender body in a stratified flow is thus a canonical case that attracts a lot of attention \cite{lin1979wakes,spedding1996turbulence,pal2017direct,ortiz2019stratified,zhou2019large}. 
The wake exhibits multiple stages along the downstream developing direction, and rich physics has been observed in this process \cite{nidhan2021coherence}. 

\begin{figure}
    \centering
    \includegraphics[width=0.8\textwidth]{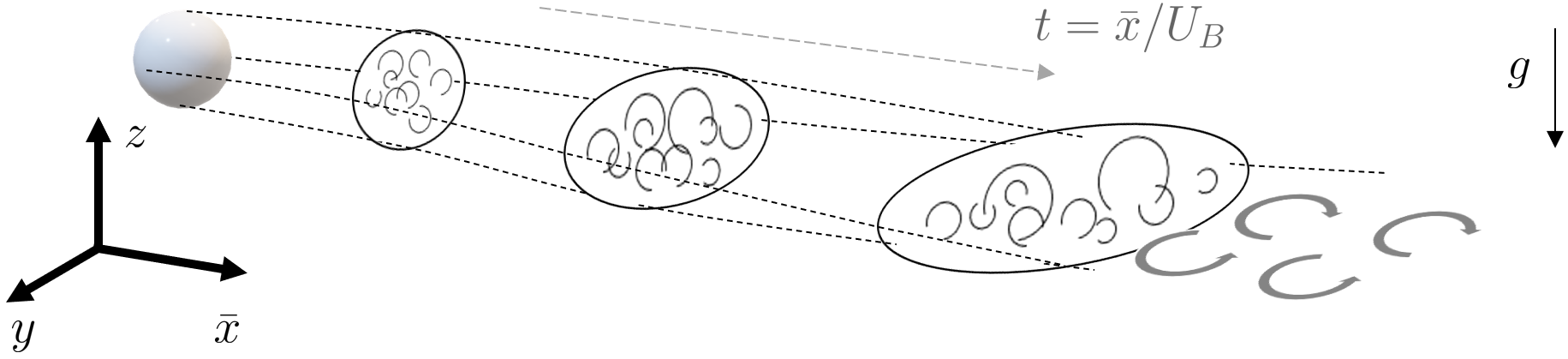}
    \caption{The sketch of a stratified wake behind a towed sphere. 
    Several cross sections of the downstream locations are drafted, and the wake core is separated from the surrounding by the solid lines. 
    The dashed lines indicate the trajectory of the wake core boundary. 
    The collapsed large-scale horizontal structures are formed at the late wake regime. 
    We use $\overbar{x}$ as the downstream location in the laboratory frame, $y$ as the spanwise direction, and $z$ for the vertical direction. 
    The gravity is in the $-z$ direction. 
    The temporal simulation is evolved in the $t=\overbar{x}/U_B$ direction. }
    \label{fig:intro-sketch}
\end{figure}

We show the sketch of the stratified wake in figure \ref{fig:intro-sketch}. 
In this paper, we examine the flow over a towed sphere in a uniformly stratified environment, where the characteristic velocity is the towing speed $U_B$, and the characteristic length is the diameter of the sphere $D$. 
The global Reynolds number is $Re_B=U_B D/\nu$, where $\nu$ is the kinematic viscosity. 
The global Froude number is $Fr_B=U_B/ND$, where the Brunt-V\"{a}is\"{a}l\"{a} frequency $\displaystyle N\equiv\sqrt{-\frac{g}{\rho_0}\frac{\partial \bar{\rho}}{\partial z}} = \sqrt{\frac{g}{T_0}\frac{\partial \bar{T}}{\partial z}}$ is the reciprocal of the buoyancy time scale, which indicates the strength of the background temperature gradient. 
Here, $\rho_0, T_0$ are the characteristic fluid density and fluid temperature, and ${\partial \bar{\rho}}/{\partial z}, {\partial \bar{T}}/{\partial z}$ are the background density gradient and the background temperature gradient. 
Therefore, $Re_B$ and $Fr_B$ are the control parameters for the stratified wake, where $Re_B$ indicates the initial ratio between the inertia and the viscous force, and $Fr_B$ indicates the initial ratio between the inertia and the buoyancy. 
However, the relative strength of the viscous force, the buoyancy, and the inertia are not constant throughout the wake. 
A local horizontal Froude number can be defined as $Fr_H=U_{0}/WN$, where $U_{0}$ is the centerline velocity deficit, and $W$ is the local wake width.
The centerline velocity deficit is decaying, while the local horizontal wake size is growing as sketched, leading to a decreasing local horizontal Froude number.
This also explains why the flow is multi-stage \cite{de2019effects,caulfield2020open}. 

Spedding \cite{spedding1997evolution} first classified these regimes into the near-wake regime (NW), the non-equilibrium regime (NEQ), and the quasi-2D regime (Q2D) via the variations in the decay rates of the centerline velocity deficit.
One of the most evident flow phenomena in all three regimes is the meandering of the turbulent wake, which refers to the large-scale, low-frequency motions observed in the wake of the bluff body \cite{yang2019review}. 
The meandering can be seen as a result of coherent structures in the turbulent flow. 
In the NW regime, wake meandering originates from the instability found near the bluff body, where both the Kelvin-Helmholtz billows and the lee waves are observed \cite{xiang2017dynamic}. 
The instability is especially evident in the high Reynolds number cases. 
The dynamics in the NW regime are similar to the unstratified counterpart since the buoyancy is relatively weak \cite{smyth2000length}. 
However, the NW regime is recognized to be short due to the growing buoyancy impact \cite{lin1979wakes}. 

Later on, the meandering starts to interact with the stratification in the NEQ regime. 
High Reynolds number cases even develop secondary instability events \cite{diamessis2011similarity}. 
Vortex structures interact with each other and are elongated in the streamwise direction after merging and straining \cite{spedding2001anisotropy}. 
According to Chongsiripinyo et al. \cite{chongsiripinyo2020decay}, the smallest eddy size that is restrained from the buoyancy overturning decreases gradually. 
Therefore, the buoyancy effect grows relatively stronger and has a greater impact on the flow statistics.  
The vertical meandering is suppressed by the buoyancy, while the flow still grows in the horizontal direction, and thus the flow grows non-axisymmetric in its mean profile, and then further impacts the isotropy of the turbulence. 

In the end, the flow enters the Q2D regime, where the flow still has strong vertical shearing but has rather insignificant vertical velocities.
Wake meandering is mainly in the horizontal direction in this regime as large horizontal structures are formed since the buoyancy effect is strong enough to fully collapse the flow in the vertical direction. 
There is little vertical transport, and vortices are decoupled into layers in the vertical direction, which leads to the formation of thin layered vortical patches \cite{basak2006dynamics,spedding1996turbulence}. 
The horizontal motions constitute the majority of the kinetic energy \cite{spedding2002vertical}. 
The emergence of the large horizontal structures is universally seen in the stratified wake, regardless of the stratification strength \cite{spedding1997evolution}.

The numerical study of stratified wakes stems from two types of simulation frameworks -- the temporally evolving frame and the spatially evolving frame.
Simulations in the spatially evolving frame include the wake generator body in the computation domain and can capture flow separation and vortex shedding from the body \cite{pal2017direct,chongsiripinyo2020decay}.
However, the high computational cost associated with resolving the boundary layer near the body and wake development far from the body makes it infeasible to capture the late wake physics.
Compared to a spatially evolving frame, simulations in the temporally evolving frame are much more cost-effective and have been utilized as the workhorse to study the late wake phenomenon \cite{brucker2010comparative,diamessis2011similarity,zhou2019large}.
The elapsed time in a temporal simulation is interchangeable with downstream distance by the relationship $\bar{x}=U_Bt$, where $\bar{x}$ is the distance behind the body in the streamwise direction and $t$ is the elapsed time in the temporally evolving simulation.
On the other hand, the initialization of temporally evolving wakes requires the wake characterization at a specific time. 
Thus, it is sensitive to the choice of initial condition in the early wake \cite{redford2012universality}, but the memory of initialization is lost in the late wake \cite{meunier2004loss}.

Although the meandering effect has been widely observed in stratified wakes, especially in the Q2D regime \cite{lin1979wakes}, it is by far not much studied.
As one of those important flow dynamics, the meandering effect is embedded throughout the flow development and becomes significant in the late wake.
Quantifying the meandering enables the detection of wake far downstream of the body in a stratified environment \cite{spedding2014wake,chinta2022regime}. 
In addition, the interaction of the meandering with other flow structures will impact the scaling of velocity and has not been examined \cite{lin1979wakes}. 
The change of the scaling behavior will be further related to our understanding of regime classification \cite{zhou2019large,chongsiripinyo2020decay}. 
On the other hand, it has been studied extensively for wind turbines in boundary layers. 
In the wake of wind turbines, the meandering effect impacts the wake recovery rate and the turbine load \cite{yang2019review}. 
Models for meandering have been developed to incorporate the meandering effect \cite{keck2014atmospheric}. 

In this paper, we study the meandering effect by isolating it from the stationary velocity, i.e., the velocity without meandering.
Temporal DNS with $\mathcal{O}(100)$ repetitions is used in order to obtain enough statistic samples.
The temporal simulation is preferred in this study because it allows us to study the meandering effect in the late wake.
The aim of this study is to clarify the effects of meandering on the self-similarity assumptions in the horizontal and vertical directions.
This will further clarify the effect of meandering on the scaling of wake statistics.
Specifically, we extract the center location, wake width, and wake height of the instantaneous flow field and examine how they develop and how the control parameters $Re_B$ and $Fr_B$ change their behavior. 
The mean velocity profile is thus the superposition of the meandering effect to the stationary velocity profile. 
The stationary profile from the DNS can be directly compared with the modeled wake profile, and the comparison further verifies the self-similar assumption. 
We will establish the theoretical basis of the observed momentum conservation in the stratified wake \cite{li2024direct}.
A similar idea of such superposition is proposed for the turbulent boundary layer \cite{krug2017revisiting,li2021experimental}, where the wake profile for a boundary layer can be modeled as a blending of the free stream and the log layer with a fluctuating interface. 
However, there are non-trivial differences when applying the blending method to the stratified wake. 
The existence of gravity leads to the destruction of the axisymmetry. 
The anisotropy results in a two-dimensional velocity profile and makes it more complicated than the one-dimensional blending in the boundary layer. 


The remainder of the paper is organized as follows. 
We first elaborate our simulation details and case setup in Sec. \ref{sec:sim}. 
In Sec. \ref{sec:model}, we explain the extraction of the wake meandering characteristics and develop the theoretical analyses on the superposition of the meandering to the stationary profile based on the two-dimensional self-similarity assumption. 
Then, we show the results of the meandering effect in Sec. \ref{sec:results}, where we examine the self-similarity assumption, the development of the meandering properties, and the embedded flow structures out of conditional averaging. 
Lastly, the paper is concluded in Sec. \ref{sec:conclusion}.

\section{Simulation details}
\label{sec:sim}

We conduct direct numerical simulations of the wake behind a towed sphere in a stably stratified flow, where the background temperature gradient is $\partial \overbar{T}/\partial z$. 
Five cases with different flow control parameters are examined for $\mathcal{O}(100)$ repetitions, and we therefore can obtain ensemble-averaged flow fields. 
Following \cite{brucker2010comparative}, we employ the temporally evolving framework in order to obtain far wake statistics, which is not available in body-inclusive simulation due to high computational cost. 
In temporal simulations, the computational domain is periodic in the streamwise direction ($x$) and the simulation evolves in time. 
By employing temporal flow approximation, the downstream distance from the sphere, $\overbar{x}$ in the laboratory frame, is interchangeable with the time in a temporal simulation, $t$, since the streamwise gradients are small compared to the lateral gradients. 
To be exact, the downstream distance $\overbar{x}$ and the time of the temporal simulation are related through $\overbar{x}=U_Bt$.
The streamwise homogeneity allows streamwise averaging, which has been utilized and verified in many previous studies \cite{gourlay2001numerical,de2012simulation,zhou2016surface,zhou2019large}.
A sponge layer with thickness $L_{\rm sponge}/D=1$ is employed in the spanwise ($y$) and the vertical ($z$) directions to absorb the emitted internal gravity waves. 
We also have periodicity in the $y$ and $z$ directions.

\begin{table}
    \caption{Details of the 440 DNS cases carried out. 
    We use `R[$Re_B$/1000]F[$Fr_B$]' to differentiate between cases with different bulk Reynolds numbers $Re_B$, and bulk Froude numbers $Fr_B$. 
    The color and the symbol type for each case are given in `Sym.'. 
    The five cases are regridded at different $t_f$, and the period of regridding is indicated by its superscript. 
    The multiple periods of each case share the same symbol. 
    Note $t_f$ is the equivalent end time of the temporal simulation to the downstream location from the body. 
    The simulations are each independently repeated for either $80$ or $100$ times, given by the column `Rep'. 
    The temporal simulations are homogeneous in the $x$ direction, and averaging can be carried out both for the streamwise direction and for multiple repetitions. }
    \label{tab:sim-details}
    \centering
\begin{ruledtabular}
\begin{tabular}{l c c c c c c c c c c c}
    \def~{\hphantom{0}}
        Case & Sym. & $Re_B$ & $Fr_B$ & $L_x$ & $L_y$ & $L_z$ & $N_x$ & $N_y$ & $N_z$ & $t_f$ & Rep. \\ [3pt]
        \colrule
        R10F02 & \includegraphics[height=\fontcharht\font`\B]{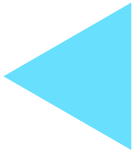} 
        & $10,000$ & $2$ & $60$ & $38$ & $14$ & $1280$ & $768$ & $256$ & $1380.18$ & $100$ \\
        R20F02$^1$ &  \includegraphics[height=\fontcharht\font`\B]{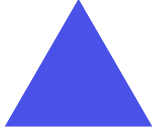} 
        & $20,000$ & $2$ & $60$ & $24$ & $12$ & $1280$ & $768$ & $384$ & $109.29$ & $100$ \\
        R20F02$^2$ & \includegraphics[height=\fontcharht\font`\B]{Intro/R20F02_sym.PNG} 
        & $20,000$ & $2$ & $60$ & $36$ & $12$ & $1280$ & $576$ & $192$ & $800.62$ & $100$ \\
        R20F10$^1$ & \includegraphics[height=\fontcharht\font`\B]{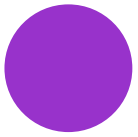} 
        & $20,000$ & $10$ & $60$ & $24$ & $12$ & $1280$ & $768$ & $384$ & $109.29$ & $80$ \\
        R20F10$^2$ & \includegraphics[height=\fontcharht\font`\B]{Intro/R20F10_sym.PNG} 
        & $20,000$ & $10$ & $60$ & $36$ & $12$ & $1280$ & $576$ & $192$ & $1220.58$ & $80$ \\
        R20F50$^1$ & \includegraphics[height=\fontcharht\font`\B]{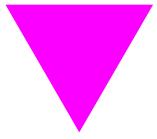} 
        & $20,000$ & $50$ & $60$ & $24$ & $12$ & $1280$ & $768$ & $384$ & $109.29$ & $80$ \\
        R20F50$^2$ & \includegraphics[height=\fontcharht\font`\B]{Intro/R20F50_sym.PNG} 
        & $20,000$ & $50$ & $60$ & $36$ & $12$ & $1280$ & $576$ & $192$ & $568.91$ & $80$ \\
        R20F50$^3$ & \includegraphics[height=\fontcharht\font`\B]{Intro/R20F50_sym.PNG} 
        & $20,000$ & $50$ & $60$ & $48$ & $24$ & $640$ & $384$ & $192$ & $5232.38$ & $80$ \\
        R50F50$^1$ & \includegraphics[height=\fontcharht\font`\B]{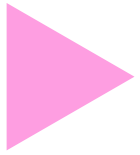} 
        & $50,000$ & $50$ & $60$ & $8.25$ & $8.25$ & $2560$ & $550$ & $550$ & $67.46$ & $80$ \\
        R50F50$^2$ & \includegraphics[height=\fontcharht\font`\B]{Intro/R50F50_sym.PNG} 
        & $50,000$ & $50$ & $60$ & $23.04$ & $11.52$ & $1280$ & $768$ & $384$ & $265.72$ & $80$ \\
        R50F50$^3$ & \includegraphics[height=\fontcharht\font`\B]{Intro/R50F50_sym.PNG} 
        & $50,000$ & $50$ & $60$ & $46.08$ & $23.04$ & $640$ & $512$ & $256$ & $5800.23$ & $80$ \\
    \end{tabular}
    \end{ruledtabular}
\end{table}

The simulations are independently and randomly initiated for all the repetitions. 
The initial profile is taken at $\overbar{x}/D=6$ behind the sphere to avoid the recirculation regime \cite{bevilaqua1978turbulence}. 
The synthesized initial velocity fields keep the same mean velocity profile and the turbulent kinetic energy measured from the experiments \cite{brucker2010comparative}, where the mean velocity profile is approximately Gaussian. 
Therefore, these initial profiles determine that the momentum flux constants $M_0=0.0275$ and $M_{s0}=0.0267$. 
The velocity fluctuations are created in the spectral space to ensure an initial broadband energy spectrum, and are exponentially damped along the radial direction. 

We use an in-house second-order finite-difference solver, which has been verified in \cite{yang2020multiple,yang2015salinity,li2021thermohaline}, to solve the incompressible Navier-Stokes equations. 
We employ the Boussinesq approximation, where the buoyancy effect is included as a body force in the momentum equations. 
The non-dimensionalized Navier-Stokes equation reads, 
\begin{equation}
    \frac{\partial{u_{i}}}{\partial{t}} + \frac{\partial{\left( u_{j} u_{i} \right)}}{\partial{x_{j}}} = -\frac{\partial{p}}{\partial{x_{i}}} + \frac{1}{Re_B} \frac{\partial^2{u_{i}}}{\partial{x_{j}} \partial{x_{j}}} - \frac{1}{Fr_B^2} {\rho'} \delta_{i3},
 \label{eq:mom}
\end{equation}
where $u_i$ is the instantaneous velocity, $i=1,2,3$ corresponds to $x,y,z$ direction, $x_i$ is the spatial coordinates, $\rho'$ is the instantaneous density fluctuation, and $\delta_{i3}$ is the Kronecker delta. 
The solver solves the transport equation of temperature:
\begin{equation}
    \frac{\partial{T}}{\partial{t}} + u_{i} \frac{\partial{(\bar{T} + T)}}{\partial{x_{i}}} = \frac{1}{Re_B Pr} \frac{\partial^2{T}}{\partial{x_{i}} \partial{x_{i}}},
 \label{eq:temp}
\end{equation}
where $Pr = \nu / \alpha$ is the Prandtl number. 
Then, the density field is obtained from a linear equation of the state
\begin{equation}
    \frac{\rho - \rho_0}{\rho_0} = -\beta(\bar{T} + T - T_0);\quad \beta = -\frac{1}{\rho_0} \left( \frac{\partial{\rho}}{\partial{T}} \right)_P,
\end{equation}
where $\bar{T}$ is the background mean temperature, $T$ is the dimensionless temperature fluctuation from the background mean value, $\rho_0$ is the reference density, $T_0$ is the reference temperature, and $\beta$ is the thermal expansion coefficient. 
All the quantities are normalized using $D$, $U_B$ and $\partial \overbar{T}/\partial z$. 
The pressure Poisson equation is solved to ensure incompressibility. 
The flow is advanced using the third-order Runge-Kutta time scheme.
More details of the code can be found in \cite{van2015pencil}, and\cite{ostilla2015multiple}.

The flow control parameters are the global Reynolds number $Re_B$ and the global Froude number $Fr_B$. 
We list the details of the five cases in Table \ref{tab:sim-details}, where ``R[$Re_B$/1000]F[$Fr_B$]'' denotes $Re_B$ and $Fr_B$. 
The high Reynolds number cases have a larger wake size and a decreasing local Reynolds number downstream, i.e., the Kolmogorov scale $\eta=(\nu^3/\varepsilon)^{1/4}$ gradually increases. 
We re-grid the flow field and enlarge the computational domain at the temporal time step $t=t_f$, when we still ensure $\Delta x_i/\eta<2\sim 4$ as required by \cite{kim1987turbulence}. 
The cases R20F02 and R20F10 have been re-gridded once, while the cases R20F50 and R50F50 have been re-gridded twice. 
Therefore, all the cases are run till the Q2D regime, which leads to a big $t_f$ for high $Re_B$ and high $Fr_B$ cases. 

Each case is independently repeated 80 or 100 times to ensure the convergence in the pre-multiplied probability density function (p.d.f.) of all the first and second-order statistics.
The averaged quantities are obtained by the combination of streamwise averaging and averaging among the independent repetitions. 
Specifically, we first average flow quantities in the streamwise direction since the streamwise homogeneity under temporal flow assumption. 
Then, the flow statistics can be obtained by further averaging among $\mathcal{O}(100)$ independent realizations due to the random initialization.

\section{Wake model with meandering}
\label{sec:model}


To understand the meandering effect, we will extract the wake center, width, and height from the instantaneous simulation data. 
Averaging the velocity profiles with the meandering effect will obtain the mean velocity profile. 
Averaging the velocity profiles without the meandering effect will obtain the mean stationary velocity profile. 
Then, we compare the mean velocity profile and the mean stationary velocity profile, and this will shed light on how meandering alters the flow statistics, which will be discussed in Sec. \ref{sec:results}. 
Next, we will develop the theory of the superposition of the meandering effect based on self-similarity assumptions. 
The modeled wake profile will be compared with the stationary profile from DNS in Sec. \ref{sec:results}, to identify possible distortion due to meandering.
To the best of the authors' knowledge, this is the first attempt to systematically quantify the meandering effect and its corresponding impact on the scaling of the statistics in the stratified wake.

\subsection{Wake center, width and height} \label{sub:model-center}

For every downstream location $\overbar{x}/D$ in the laboratory frame, the instantaneous velocity deficit is $u(y,z,t)$, where $y$ and $z$ indicate the spanwise and the vertical direction. 
Here, $t$ is the equivalent time in the temporal simulation for the downstream location $\overbar{x}=U_B t$, and the nondimensional time $Nt=Fr_B^{-1}\cdot\overbar{x}/D$. 
We illustrate this coordinate conversion in figure \ref{fig:intro-sketch}. 
Due to the stratification, the spreading of the wake size is not of the same rate in the $y$ and $z$ directions. 
To characterize the wake size, we need to find both the center location $y_c,z_c$ and the width and height $W, H$, where we use the subscript $c$ to indicate the center. 

We follow a generalized way to identify the wake lengths $W, H$, by employing the second central spatial moment of the squared instantaneous velocity deficit \cite{brucker2010comparative}. 
The center location for $u(y,z,t)$ is therefore defined by 
\begin{equation}
    y_c(t) = \frac{\int_A yu(y,z,t)^2 {\rm d} A}{\int_A u(y,z,t)^2 {\rm d} A},~~~
    z_c(t) = \frac{\int_A zu(y,z,t)^2 {\rm d} A}{\int_A u(y,z,t)^2 {\rm d} A}. 
    \label{eqn:model-center}
\end{equation}
The width and the height are computed by 
\begin{equation}
    W(t)=\sqrt{C_0\frac{\int_A (y-y_c)^2u(y,z,t)^2dA}{\int_A u(y,z,t)^2dA}}, ~~~
    H(t)=\sqrt{C_0\frac{\int_A (z-z_c)^2u(y,z,t)^2dA}{\int_A u(y,z,t)^2dA}}. 
    \label{eqn:model-length}
\end{equation}
Here, both of them are integrated through the cross-section $A$ which excludes the sponge layer applied in the simulations. 
We use $C_0$ to guarantee consistency with the diameter definition. 
The initial mean profile is Gaussian, and we find that $C_0=2$ will guarantee $W(0)=H(0)=D/2$ during the initialization. 

Therefore, we are able to extract the center location, the width, and the height for all instantaneous snapshots however chaotic they are. 
To quantify the meandering behavior, the center location, the width and the height are extracted for all instantaneous streamwise cross sections for all the repetitions.
We will study the joint p.d.f. of the wake center location $p_c(y_c,z_c,t)$ and it requires a huge amount of data to converge.
The kernel density estimation is used to obtain p.d.f. from large data samples \cite{botev2010kernel}.

The same technique can be applied to any other profiles we are interested in, as long as we substitute the instantaneous velocity deficit $u(y,z,t)$ by the corresponding quantities. 
For example, we may calculate the center location, the width, and the height for the mean velocity profile $\overbar{U}(y,z,t)=\langle u(y,z,t)\rangle$, where $\langle \cdot\rangle$ represents the ensemble average of the quantity. 
According to the symmetry, we should have $y_c=0$ and $z_c=0$ for $\overbar{U}(y,z,t)$. 
However, note that the width and the height of the mean profile are not necessarily the same as the mean of the instantaneous width and height. 
Similarly, we can also compute the width and the height of the p.d.f. $p_c$, and they indicate the meandering range of the wake center location.

\subsection{Superposition of the meandering effect} \label{sub:model-sp}

All instantaneous velocity profiles can be seen as the superposition of the meandering effect and a stationary velocity profile. 
The meandering effect is the deviation of the instantaneous center location from the mean wake center, while the stationary velocity is the velocity without meandering, and thus, its center is located at the mean wake center. 

In this section, we will conduct a statistical analysis of the meandering effect, and no presumptions about the shape of the velocity profile are made at the current stage. 
The symmetry of the sphere wake guarantees the mean wake center at the domain center through the wake development, which helps simplify the expressions, though it is not required in the analysis. 
For any instantaneous velocity profile centered at $(y_c,z_c)$, the stationary velocity profile is defined as $u_s(y,z,t)=u(y+y_c,z+z_c,t)$. 
We are able to find the mean stationary profile by 
\begin{equation}
    \overbar{U}_s(y,z,t)=\langle u_s(y,z,t)\rangle=\langle u(y+y_c,z+z_c,t)\rangle. 
    \label{eqn:model-U-stationary}
\end{equation}
We are also able to obtain the conditional stationary profile with the center location $(y_c,z_c)$ by conditional averaging, i.e., 
\begin{equation}
    U_s(y,z,y_{c},z_{c},t)=\langle u_s(y,z,t)\rangle |_{|y_{c0}-y_c|/D<\Delta_s,|z_{c0}-z_c|/D<\Delta_s}. 
    \label{eqn:model-U-conditional}
\end{equation}
Here, $\Delta_s$ is the averaging window, and the instantaneous snapshots with the center location $(y_{c0},z_{c0})$ falling in the averaging window will be used for calculation. 
We choose $\Delta_s=0.1$ to guarantee the convergence of conditional averaging. 
Conditional stationary profiles further disclose the changes in the velocity profile depending on the center location of the meandering effect. 

The mean stationary profile $\overbar{U}_s(y,z,t)$ is the collapse of the conditional stationary profile $U_s(y,z,y_c,z_c,t)$ by averaging out for various center locations. 
Correspondingly, the mean velocity profile with meandering $\overbar{U}(y,z,t)$ is therefore the collapse of the meandered conditional profile by averaging out for various center locations. 
To be exact, 
\begin{align}
    \overbar{U}_s(y,z,t)&=\iint U_s(y,z,y_c,z_c,t)p_c(y_c,z_c,t){\rm d}y_c{\rm d}z_c, \label{eqn:model-Us}\\
    \overbar{U}(y,z,t)&=\iint U_s(y-y_c,z-z_c,y_c,z_c,t)p_c(y_c,z_c,t){\rm d}y_c{\rm d}z_c.
    \label{eqn:model-superposition}
\end{align}
Here, $p_c(y_c,z_c,t)$ is the joint probability density function (p.d.f.) of the center location. 
These equations can be obtained through the definitions of the conditional average and the ensemble average \cite{antonia1981conditional}. 
The mean velocity profile and the mean stationary profile are thus bridged through the conditional stationary profile. 
They differ in whether the center location is meandering during averaging.

\subsection{Self-similarity assumptions and simplifications} \label{sub:model-ss}

Next, we involve self-similarity assumptions and formalize equations with width and height. 
The superposition equations of the meandering can thus be simplified, which helps spot the meandering effect in the scaling of the statistics. 

If we assume that the self-similarity for the mean velocity profile preserves till the late wake, we have a universal expression for the velocity profile after a coordinate transformation, 
\begin{equation}
    \overbar{U}(y,z,t) = \hat{U}(\xi,\eta,t) = U_{0}(t)f(\xi,\eta), {\rm ~~~ where~} \xi=y/W,~ \eta=z/H. 
    \label{eqn:model-ss-Udef}
\end{equation}
We use $\hat{U}$ to denote the velocity in the coordinates normalized with width and height, the subscript $0$ to denote the center velocity deficit, and $W$ and $H$ are the width and height for $\overbar{U}(y,z,t)$ calculated as in Sec. \ref{sub:model-center}. 
The shape of the velocity profile is thus invariant with respect to time $t$, and is represented by a universal function $f(\xi,\eta)$.

To separate the meandering effect from the stationary profile, we further assume self-similarity in the conditional stationary profiles $U_s(y,z,y_c,z_c,t)$ and the joint p.d.f. of the center location $p_c(y_c,z_c,t)$ throughout the downstream locations. 
The conditional stationary profiles are similarly normalized in coordinates, and the self-similarity is expressed as follows, 
\begin{equation}
\begin{split}
    U_s(y,z,y_c,z_c,t) &= \hat{U}_s(\xi_s,\eta_s,\xi_{sc},\eta_{sc},t) = U_{s0}(t)g(\xi_s,\eta_s), \\ {\rm ~~~ where~} &\xi_{s}=y/W_s,~ \eta_{s}=z/H_s,~ \xi_{sc}=y_c/W_s,~ \eta_{sc}=z_c/H_s.
\end{split}
    \label{eqn:model-ss-Usdef}
\end{equation}
The subscript $s$ represents the stationary velocity, and $W_s$ and $H_s$ are the width and height of the stationary profile. 
Similarly, $U_{s0}(t)$ is the center stationary velocity deficit, and the time-invariant stationary velocity profile shape is represented by $g(\xi_s,\eta_s)$. 

As for the joint p.d.f. of the center location, the self-similarity is expressed as follows, 
\begin{equation}
    p_c(y_c,z_c,t)=\hat{p}_c(\xi_{pc},\eta_{pc})p_{c0}(t), {\rm ~~~ where~} \xi_{pc}=y_{c}/W_{p},~ \eta_{pc}=z_{c}/H_{p},
    \label{eqn:model-ss-pdf}
\end{equation}
where the subscript $p$ represents the p.d.f. of the center location, and $p_{c0}(t)$ is the normalization term such that the integration of the joint p.d.f. across the wake cross-section is $1$ before and after the coordinate transformation, i.e.,
\begin{equation}
    \iint \hat{p}_c(\xi_{pc},\eta_{pc}){\rm d}\xi_{pc}{\rm d}\eta_{pc} = 1. 
    \label{eqn:model-pdf-int1}
\end{equation}

Now, we can express the velocity deficit $\overbar{U}(y,z,t)$ as the combination of the velocity deficit amplitude $U_{s0}(t)$ for the conditional stationary profile and the self-similar profiles. 
The simplification of the superposition equation \eqref{eqn:model-superposition} is 
\begin{equation}
    \overbar{U}(y,z,t)=U_{s0}(t)\iint g(\xi_s-\xi_{sc},\eta_s-\eta_{sc})\hat{p}_c(\xi_{pc},\eta_{pc}){\rm d}\xi_{pc}{\rm d}\eta_{pc}. 
    \label{eqn:model-ss-Uint}
\end{equation}
Note that this does not necessarily require the same width and height of the p.d.f. as those of the mean velocity profile. 
The varying of the width and the height will lead to the varying integral value, making the integral a function of $t$. 

Thus, we are able to link the centerline deficit velocities $U_{0}(t)$ and $U_{s0}(t)$ by 
\begin{equation}
    U_{0}(t) = U_{s0}(t)\iint g(\xi_{sc},\eta_{sc})\hat{p}_c(\xi_{pc},\eta_{pc}){\rm d}\xi_{pc}{\rm d}\eta_{pc}. 
    \label{eqn:model-ss-U0superpose}
\end{equation}
It always stands that $U_{0}(t)\leq U_{s0}(t)$ since $g\leq 1$, so the meandering effect will reduce the mean velocity deficit. 

The centerline velocity deficit therefore collapses to be 
\begin{equation}
\begin{split}
    U_{0}(t) &= C_s(t) U_{s0}(t), \\ 
    {\rm ~~ where ~} C_s(t) & = \iint g(\xi_{sc},\eta_{sc})\hat{p}_c(\xi_{pc},\eta_{pc}){\rm d}\xi_{pc}{\rm d}\eta_{pc}. 
\end{split}
    \label{eqn:model-ss-U0superpose-const}
\end{equation}
If the proportional relationship between $W_s, H_s$ and $W_{p},H_{p}$ keeps constant as flow develops, we will obtain $C_s(t)={\rm const}$. 
Therefore, the meandering will lead to a constant reduction in the mean velocity deficit $U_0$. 
Otherwise, when the proportional relationship between the widths and the heights changes, the scaling of the velocity profile will be impacted. 

Intuitively, $p_c(y_c,z_c,t)$, as the possibility of the meandering location $(y_c,z_c)$, is the quantification of the meandering behavior. 
We know that the center location $y_{cp}=0$ and $z_{cp}=0$ due to the symmetry. 
We can also extract the width and height of the p.d.f., $W_{p}, H_{p}$. 
They are the measurement of the range of the meandering in the horizontal and the vertical directions. 
A large $W_{p}$ is expected in the late wake regime, since the large-scale structures will be formed in the stratified environment through vortex interactions and the instantaneous horizontal center location is expected to vary greatly. 

Up till now, only self-similarity is assumed. 
If we further assume that $g$ and $\hat{p}_c$ are standard 2D Gaussian distributions, the convolution will bring that $f$ is also 2D Gaussian distributed, and that the superposed width and height follow
\begin{equation}
    W_{sp}^2=W_s^2+W_p^2,~~ H_{sp}^2=H_s^2+H_p^2. 
    \label{eqn:model-ss-Gauss-length}
\end{equation}

To evaluate the variation in the instantaneous profiles, we may also study the p.d.f. of the wake width and height $p_l(W,H,t)$, where the subscript $l$ indicates the length scales, i.e., the width and the height. 
Intuitively, $p_l(W,H,t)$, as the possibility of a specific pair of $(W, H)$, indicates the possible distortion of the instantaneous velocity. 
If we have a large span of $p_l(W,H,t)$ in space, the instantaneous velocity varies strongly in its profile. 
Overall, both joint p.d.f. profiles, $p_c(y_c,z_c,t)$ and $p_l(W,H,t)$, are the quantification of the meandering behavior.

\subsection{Momentum flux equation} \label{sub:model-mom}

Here, we would like to build the connection between the scaling of the velocity and the scaling of the width and the height based on the self-similarity assumption. 
We start with the momentum conservation law \cite{pope2000turbulent,tennekes1972first}, which says 
\begin{equation}
    \iint \rho \overbar{U}(U_B-\overbar{U}){\rm d}y{\rm d}z={\rm const}, 
    \label{eqn:model-momcons}
\end{equation}
where $U_B$ is the tow speed, i.e., the free stream velocity. 
Due to the temporal flow approximation, the momentum conservation law stands along the time direction for temporal simulation. 
If we only consider the flow far downstream that the velocity deficit is weak compared to the tow speed, i.e., $\overbar{U}\ll U_B$, we are able to simplify the momentum conservation law to 
\begin{equation}
    \iint \overbar{U}(y,z,t){\rm d}y{\rm d}z={\rm const}.
    \label{eqn:model-momcons-linear}
\end{equation}
The constant indicates the streamwise momentum flux, and it remains constant along time direction. 
Based on the self-similarity assumption of the mean velocity, we may therefore reduce equation \eqref{eqn:model-momcons-linear} to 
\begin{equation}
    \iint \overbar{U}(y,z,t){\rm d}y{\rm d}z
    =U_{0}(t)W(t)H(t) \iint f(\xi,\eta){\rm d}\xi{\rm d}\eta 
    ={\rm const},
    \label{eqn:model-ss-momcons}
\end{equation}
i.e., 
\begin{equation}
    U_{0}(t)W(t)H(t)  =M(t) U_B D^2, 
    \label{eqn:model-ss-momcons2}
\end{equation}
where $M(t)=M_0$ shall be a nondimensional constant for the conservation law of the momentum flux. 
Note that the simplified conservation law does not require the exact expression of the self-similar profile, and thus Gaussian profile is not required in this scenario. 
It stands as long as self-similarity exists for the velocity profile with a weak velocity deficit.

Similarly, we apply the momentum conservation to the mean stationary profile $\overbar{U}_s(y,z,t)$ based on the self-similarity equation \eqref{eqn:model-ss-Usdef} and obtain the same simplified equation for $U_{s0}$ and $W_s,H_s$, 
\begin{equation}
    U_{s0}(t)W_s(t)H_s(t)  =M_s(t) U_B D^2,
    \label{eqn:model-ss-momstation}
\end{equation}
where $M_s(t)=M_{s0}$ shall also be a constant through time according to the conservation law. 

We know that the flow deficit decays fast downstream. 
Exceptions are mainly in the near wake regime for low $Fr$ case \cite{pal2017direct}. 
Therefore, the conservation of the momentum flux is guaranteed in the late wake when the self-similarity assumptions stand, and this will be verified in Sec. \ref{sub:results-momcons}.

\section{Results}
\label{sec:results}

\subsection{Visualization} \label{sub:results-inst}

\begin{figure}
    \centering
    \includegraphics[width=0.63\textwidth,valign=t]{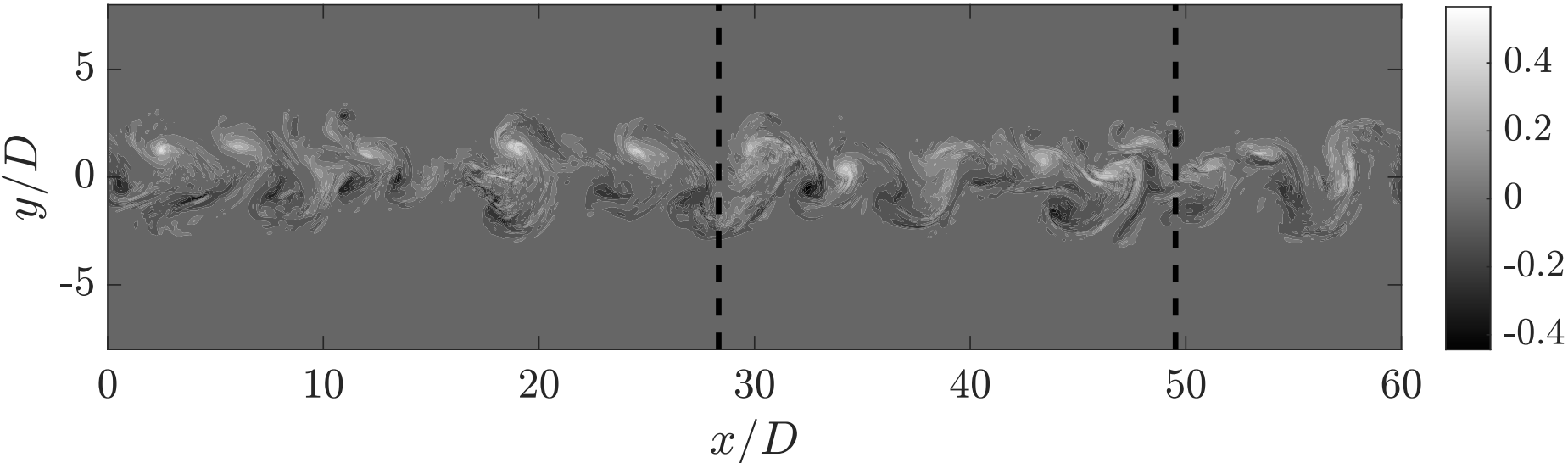}
    \hspace{-0.64\textwidth}(a)\hspace{0.6\textwidth}~\\
    \vspace{2mm}
    \includegraphics[height=0.16\textwidth,valign=t]{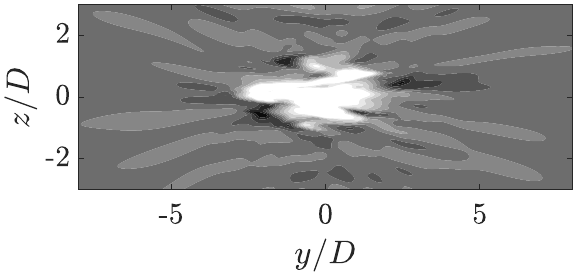}
    \hspace{-0.36\textwidth}(b)\hspace{0.33\textwidth}
    ~(c)\hspace{1mm}\includegraphics[height=0.16\textwidth,valign=t]{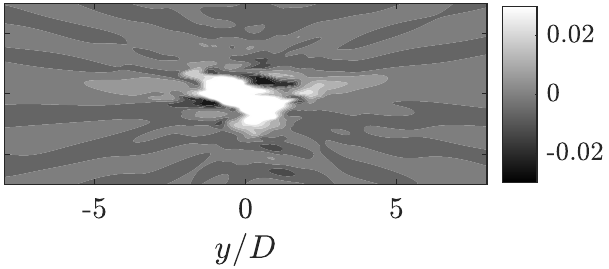}
    \caption{The instantaneous flow fields for R20F02 at $Nt=45.6$, which falls in the NEQ regime. 
    (a) The normalized vertical vorticity $\omega_z D/U_B$ in the $x-y$ plane at the wake center. 
    (b,c) The normalized streamwise velocity $u/U_B$ in the $y-z$ plane extracted at (b) $x/D=28.3$ and (c) $x/D=49.5$, denoted by the black dashed lines in (a). 
    The visualization is limited to $-8<y/D<8$ and $-3<z/D<3$ for clarity. }
    \label{fig:results-instant-1}
\end{figure}

\begin{figure}
    \centering
    \includegraphics[width=0.63\textwidth,valign=t]{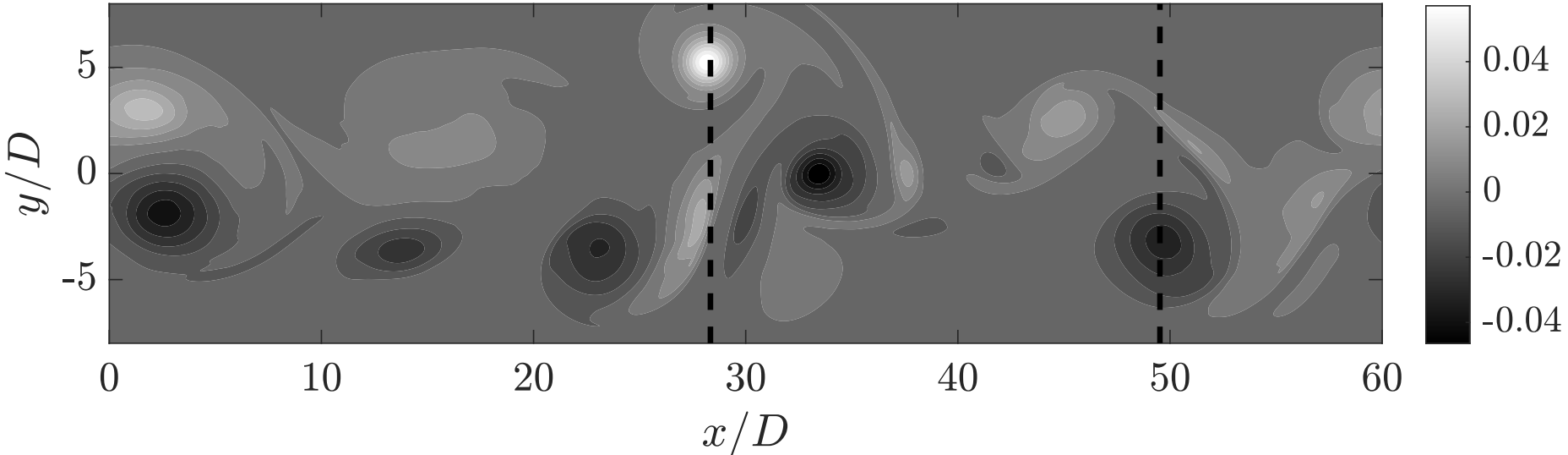} 
    \hspace{-0.64\textwidth}(a)\hspace{0.6\textwidth}~\\
    \vspace{2mm}
    \includegraphics[height=0.16\textwidth,valign=t]{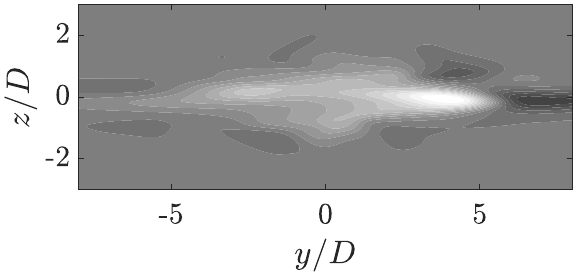}
    \hspace{-0.36\textwidth}(b)\hspace{0.33\textwidth}
    ~(c)\hspace{1mm}\includegraphics[height=0.16\textwidth,valign=t]{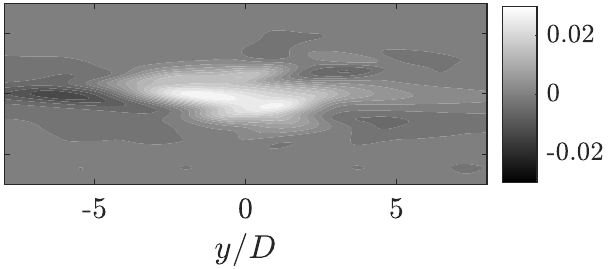}
    \caption{The instantaneous flow fields for R20F02 at $Nt=400.3$, which falls in the Q2D regime. 
    The contour settings are the same as figure \ref{fig:results-instant-1}. }
    \label{fig:results-instant-2}
\end{figure}

We first visualize the meandering of a stratified turbulent wake in the instantaneous flow fields. 
Figures \ref{fig:results-instant-1} and \ref{fig:results-instant-2} are the normalized instantaneous vertical vorticity $\omega_z D/U_B$ in the horizontal $x-y$ plane at the wake vertical center, with $Nt=45.6$ and $Nt=400.3$. 
They are both from the case R20F02, and they show typical vortical structures in the NEQ regime and in the Q2D regime respectively. 
In the NEQ regime as shown in figure \ref{fig:results-instant-1} (a), asymmetric structures with vertical vorticity $\omega_z$ of opposite signs are densely packed along the centerline, and shed downstream. 
The turbulent flow in the wake core is composed of fine turbulent eddies of multiple scales. 
In figure \ref{fig:results-instant-1} (b,c), We pick two $x/D$ locations where we observe strong meandering behavior along the $y$ direction, and show the corresponding instantaneous streamwise velocity $u/U_B$ in the $y-z$ planes. 
At both locations, the wake core is more stretched in the horizontal direction than in the vertical direction because of the stratification. 
Also, the meandering behavior is stronger in the horizontal direction as the wake center is observed to be more deviated from the mean wake center. 
In the Q2D regime in figure \ref{fig:results-instant-2} (a), the vortical structures are larger in the horizontal direction and they are more isolated from each other due to the pairing and merging of the wake cores \cite{spedding2002streamwise,basak2006dynamics}. 
Fine-scale structures are less visible in the turbulent flow compared to the NEQ regime in figure \ref{fig:results-instant-1} (a), and there are no internal gravity waves as in figure \ref{fig:results-instant-1} (b,c). 
In the meantime, the horizontal meandering grows larger with the growth of the wake width, while the vertical meandering is still relatively minimal as shown in figure \ref{fig:results-instant-2} (b,c).

\begin{figure}
    \centering
    \includegraphics[height=0.18\textwidth,valign=t]{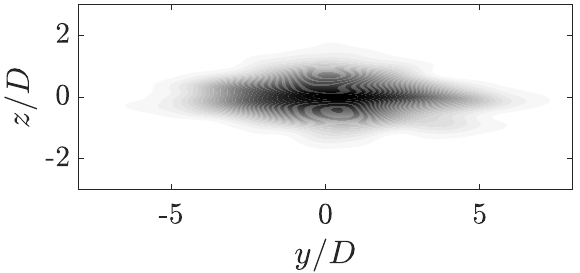}
    \hspace{-0.4\textwidth}(a)\hspace{0.36\textwidth}
    ~~~\includegraphics[height=0.18\textwidth,valign=t]{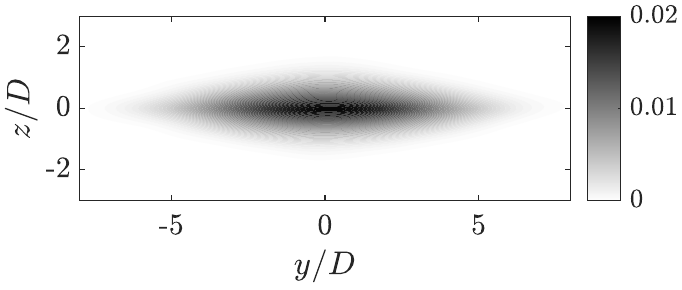}
    \hspace{-0.44\textwidth}(b)\hspace{0.4\textwidth}~
    \caption{The comparison of the mean velocity in the $y-z$ plane between (a) the 100-repetition ensemble average and (b) the 1-repetition ensemble average. 
    They are both from the case R20F02 at $Nt=400.3$. }
    \label{fig:results-ensemble}
\end{figure}

We obtain the statistics through averaging.
In temporal simulations, the statistics are traditionally obtained through the streamwise average due to the homogeneity in the streamwise direction \cite{gourlay2001numerical,de2012simulation,rowe2020internal}. 
However, only using the streamwise average is not enough to obtain a converged velocity profile as is shown in figure \ref{fig:results-ensemble} (a), which falls in the Q2D regime and we may exclude the influence of the internal gravity. 
The boundary of the wake core is oscillatory, which indicates the insufficiency of the sampling when the averaging is from only 1 repetition. 
In our calculation, we initialize each case independently 80 or 100 times, and the average is the average both in the streamwise direction and among the independent repetitions. 
The 100-repetition ensemble average in figure \ref{fig:results-ensemble} (b) shows better convergence, and the improvement happens to other time steps and other cases we work with, especially in the Q2D regime where the horizontal meandering is large.
A detailed discussion on the convergence error of the ensemble average can be found in \cite{li2024direct}.

\subsection{Two-dimensional self-similarity} \label{sub:results-2dss}

\begin{figure}
    \centering
    \includegraphics[height=0.16\textwidth]{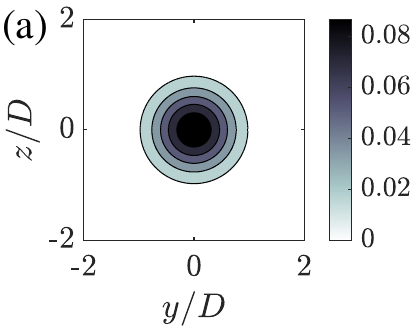}
    \includegraphics[height=0.16\textwidth]{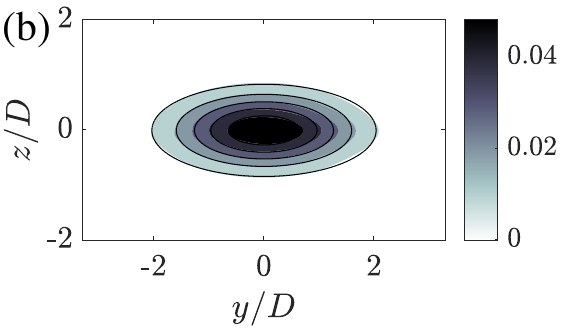}
    \includegraphics[height=0.17\textwidth]{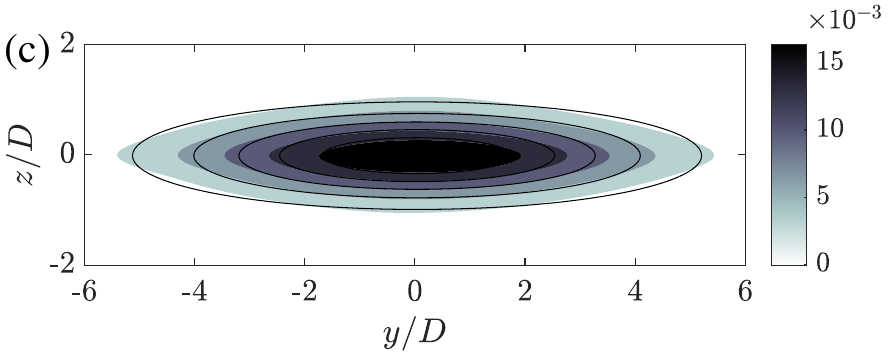}
    \includegraphics[height=0.155\textwidth]{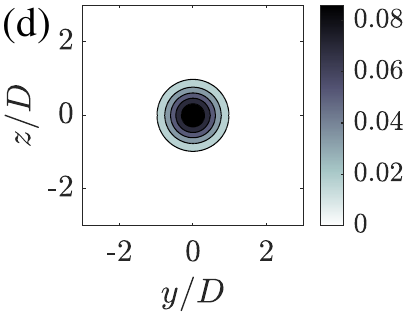}
    \includegraphics[height=0.17\textwidth]{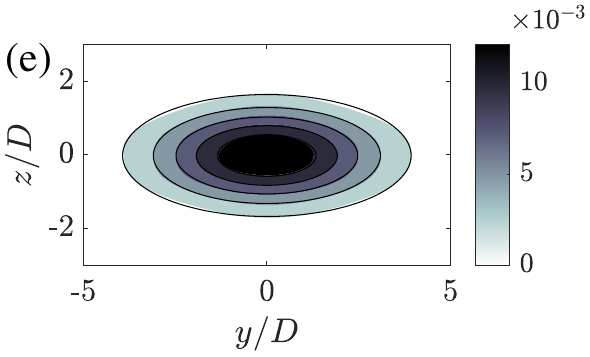}
    \includegraphics[height=0.17\textwidth]{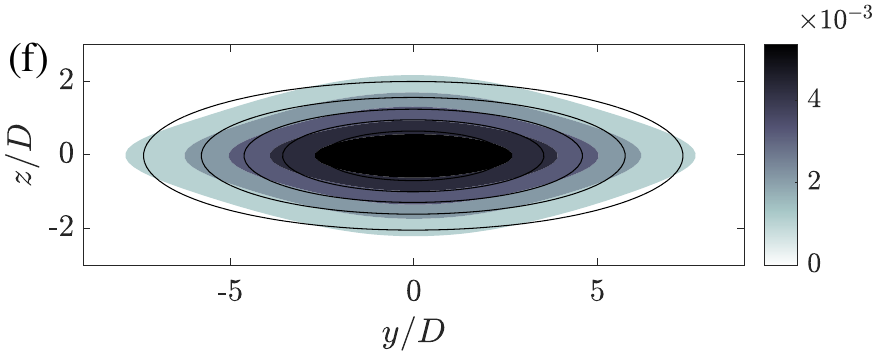}
    \caption{The mean velocity field $\overbar{U}(y,z,t)$ in the $y-z$ plane for (a,b,c) R20F02 and (d,e,f) R20F50. 
    The solid lines are the same contour levels from the 2D Gaussian fit of $\overbar{U}$. 
    The examined time steps are (a) $Nt=4.1$, (b) $Nt=45.6$, (c) $Nt=400.3$ in R20F02 and (d) $Nt=0.2$, (e) $Nt=18.6$, (f) $Nt=104.6$ in R20F50. 
    Note that the contours have different axis limits due to the change in the wake core size. }
    \label{fig:results-gauss-origin}
\end{figure}

We now examine the self-similarity assumption of the mean velocity profiles $\overbar{U}(y,z,t)$. 
As the initial mean velocity profile is generated to be Gaussian based on experiments, the self-similarity assumption indicates that the mean velocity profile should continue to be Gaussian as the flow evolves downstream. 
However, the existence of stratification changes the 2D Gaussian shape in mean velocity profiles, and we are interested in what criteria can be used to identify such change. 
The self-similarity assumption was usually examined in one-dimension along $y$- and $z$-centerlines \cite{spedding1996turbulence}. 
The normalized 1D profiles are close to the Gaussian function, and the profile along $z$-centerline has a thicker tail than the profile along $y$-centerline in the late wake \cite{spedding2001anisotropy}, which indicates the change in the shape of the mean velocity profiles. 
The anisotropy in the $y$- and $z$-directions further indicates the possibility of such distortion. 
In figure \ref{fig:results-gauss-origin}, we compare the mean velocity field in the $y-z$ plane, indicated by the filled contours, and their 2D least-square Gaussian fit, indicated by the solid contour lines, for all three regimes in two cases with different $Fr_B$, R20F02 and R20F50. 
The contours of both cases R20F02 and R20F50 are expanding in the horizontal $y$-direction and are suppressed in the vertical $z$-direction. 
The height of the mean velocity profile is smaller for case R20F02 in figure \ref{fig:results-gauss-origin} (c) when the width is comparable with R20F50 in figure \ref{fig:results-gauss-origin} (f). 
Smaller $Fr_B$ in R20F02 means a stronger background stratification, and leads to a larger ratio between width and height $W/H$ in the Q2D regime. 
The velocity profiles in both the NW and the NEQ regimes are close to Gaussian behavior. 
However, we qualitatively see that the velocity profiles in the Q2D regime for both cases do not collapse with the 2D Gaussian fit.
To conclude, we observe the distortion in the mean velocity profile as the flow evolves. 
It is not related to the ratio between width and height $W/H$, but more related to the flow transitioning to another regime.

\begin{figure}
    \centering
    \includegraphics[height=0.18\textwidth,valign=t]{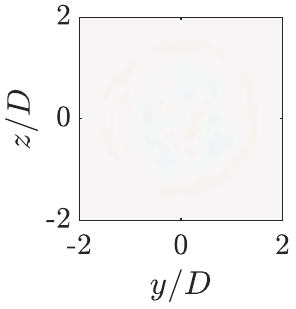}
    \hspace*{-0.18\textwidth}(a)\hspace*{0.16\textwidth}
    \includegraphics[height=0.18\textwidth,valign=t]{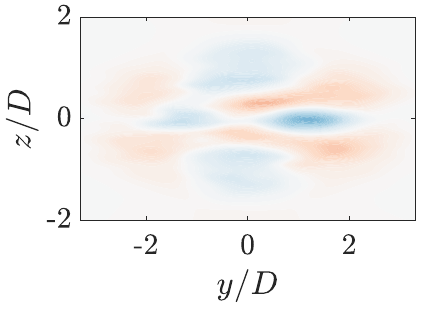}
    \hspace*{-0.26\textwidth}(b)\hspace*{0.26\textwidth}
    \includegraphics[height=0.18\textwidth,valign=t]{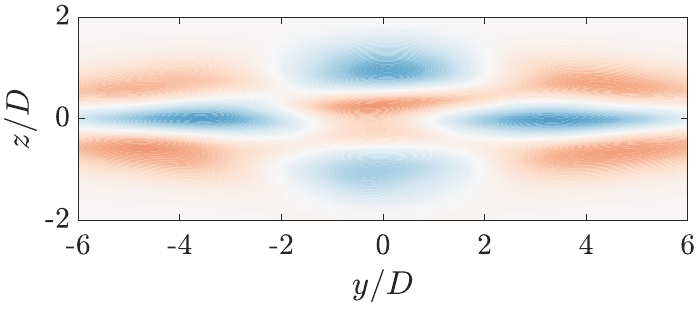}
    \hspace*{-0.41\textwidth}(c)\hspace*{0.40\textwidth}~\\
    \includegraphics[height=0.172\textwidth,valign=t]{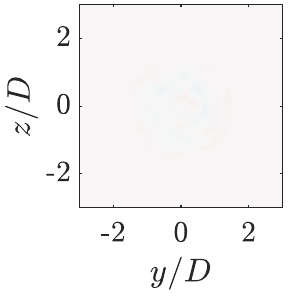}
    \hspace*{-0.18\textwidth}(d)\hspace*{0.16\textwidth}
    \includegraphics[height=0.172\textwidth,valign=t]{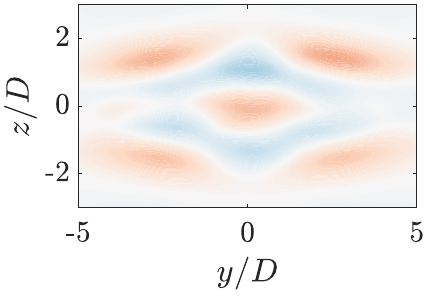}
    \hspace*{-0.26\textwidth}(e)\hspace*{0.26\textwidth}
    \includegraphics[height=0.172\textwidth,valign=t]{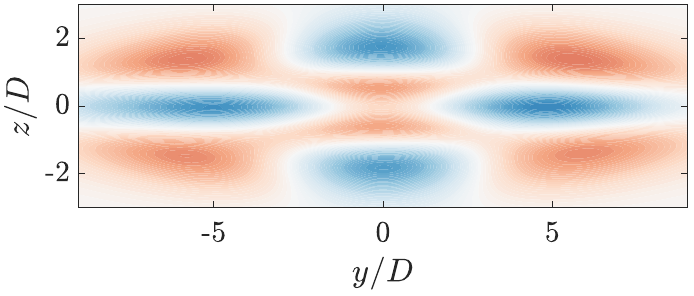} 
    \hspace*{-0.41\textwidth}(f)\hspace*{0.40\textwidth}~\\
    \vspace{2mm}
    \includegraphics[height=0.042\textwidth]{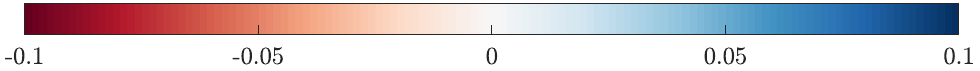}
    \caption{The difference between $\overbar{U}(y,z,t)$ and its 2D Gaussian fit for the same time steps and cases as in figure \ref{fig:results-gauss-origin}. 
    They fall respectively in the NW regime, the NEQ regime and the Q2D regime from left to right. 
    All the profiles are normalized by the maximum value $U_{0}$ of the 2D Gaussian fit. }
    \label{fig:results-gauss-origin-diff}
\end{figure}

\begin{figure}
    \centering
    \includegraphics[width=0.40\textwidth]{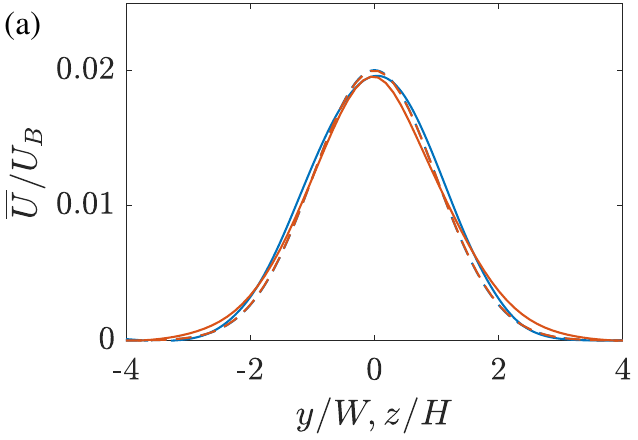}~~~
    \includegraphics[width=0.36\textwidth]{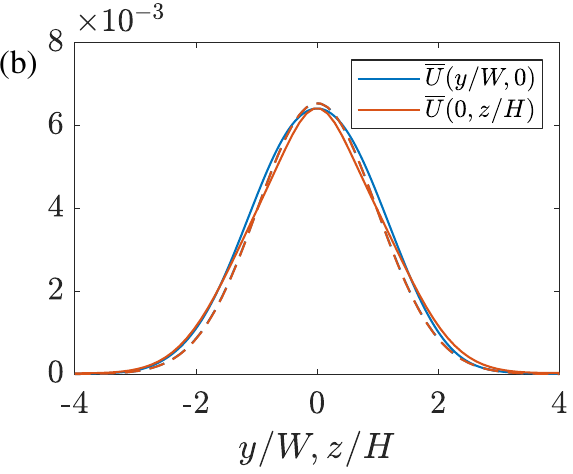}
    \caption{The velocity profiles $\overbar{U}$ along the horizontal centerline $\overbar{U}(y/W,0)$ and the vertical centerline $\overbar{U}(0,z/H)$. 
    The examined time steps are (a) $Nt = 400.3$ in R20F02 and (b) $Nt=104.6$ in R20F50, both of which lie in the Q2D regime. 
    We use the dashed lines to represent the extracted centerline profiles of the 2D Gaussian fit. }
    \label{fig:results-gauss-origin-marg}
\end{figure}

To quantitatively study the deviation of the mean velocity profile from its 2D Gaussian fit, we further examine the normalized difference between the mean velocity profile $\overbar{U}(y,z,t)$ and its 2D Gaussian fit in figure \ref{fig:results-gauss-origin-diff}. 
They are again examined for all three regimes in two cases with different $Fr_B$, R20F02 and R20F50. 
The red indicates that the $\overbar{U}(y,z,t)$ is lower than its 2D Gaussian fit, while the blue indicates that the $\overbar{U}(y,z,t)$ is higher than its 2D Gaussian fit. 
The larger the absolute value is, the more deviated $\overbar{U}(y,z,t)$ is from its 2D Gaussian fit. 
Here, we define a difference of 5.0\% as a large deviation. 
We observe that the normalized deviation from the Gaussian profile is negligible in the NW regime in figure \ref{fig:results-gauss-origin-diff} (a,d), and gradually grows to be relatively large in the Q2D regime in figure \ref{fig:results-gauss-origin-diff} (c,f).
In the NW regime and the NEQ regime, as in figure \ref{fig:results-gauss-origin-diff} (a,b,d,e), the deviations are small enough in the whole domain, the maximum absolute of which is 4.7\%. 
In the Q2D regime, the deviation is much larger. 
As in figure \ref{fig:results-gauss-origin-diff} (c,f), the maximum deviation from the Gaussian profile reaches 6.5\%. 
The normalized areas of the large deviation, i.e., the absolute deviation is more than 5.0\%, are 0.58$D^2$ and 4.50$D^2$ in figures \ref{fig:results-gauss-origin-diff} (c,f) respectively, and these are comparable to the size of the whole wake core. 

The pattern of the deviation is further examined to evaluate the possibility of a simple fix to obtain a modified self-similar profile. 
To do this, we present in figure \ref{fig:results-gauss-origin-marg} the streamwise velocity profiles $\overbar{U}$ and their 2D Gaussian fit along the horizontal centerline $\overbar{U}(y/W,0)$ and the vertical centerline $\overbar{U}(0,z/H)$ in the Q2D regime for both R20F02 and R20F50 cases. 
We observe that the profiles along both the horizontal centerline and the vertical centerline are a bit wider than the Gaussian profile. 
This observation is consistent with the positive areas near the centerlines of the contours in figure \ref{fig:results-gauss-origin-diff}. 
Away from the centerlines, the areas are mainly negative in figure \ref{fig:results-gauss-origin-diff}, and this indicates thinner profiles. 
Nonetheless, the profiles along the vertical centerline in \ref{fig:results-gauss-origin-marg} have a thicker tail than the 2D Gaussian fit, but the profiles along the horizontal centerline do not, which is also observed in \cite{spedding2001anisotropy}. 
This shows a clear difference between the horizontal direction and the vertical direction. 
The difference between the area near centerlines and the area away from centerlines, and the difference between the horizontal direction and the vertical direction deny the possibility of a simple modification to the self-similar profile after local normalization. 
The self-similarity is distorted by the existence of stratification, and the distortion grows stronger as the flow evolves.

\begin{figure}
    \centering
    \includegraphics[height=0.18\textwidth,valign=t]{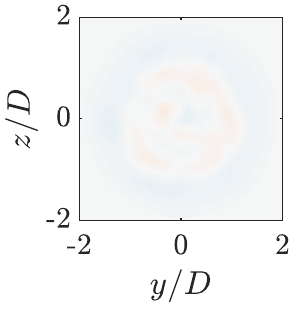}
    \hspace*{-0.18\textwidth}(a)\hspace*{0.16\textwidth}
    \includegraphics[height=0.18\textwidth,valign=t]{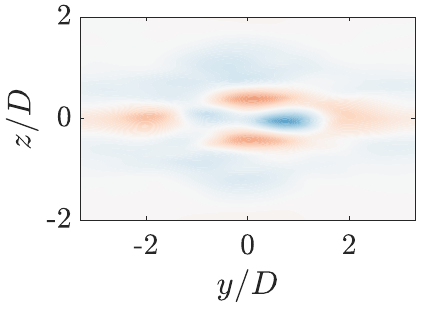}
    \hspace*{-0.26\textwidth}(b)\hspace*{0.26\textwidth}
    \includegraphics[height=0.18\textwidth,valign=t]{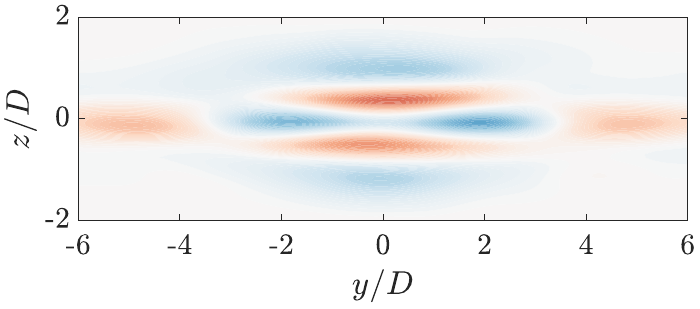}
    \hspace*{-0.41\textwidth}(c)\hspace*{0.42\textwidth}~\\
    \includegraphics[height=0.172\textwidth,valign=t]{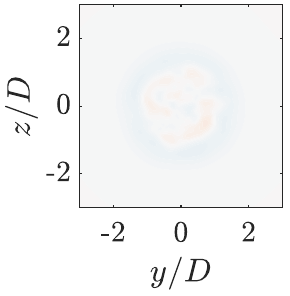}
    \hspace*{-0.18\textwidth}(d)\hspace*{0.16\textwidth}
    \includegraphics[height=0.172\textwidth,valign=t]{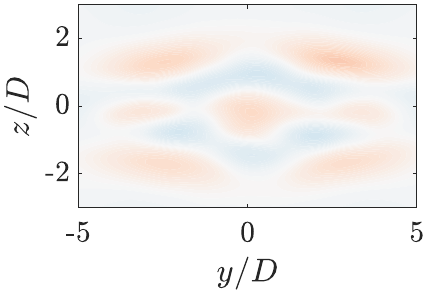}
    \hspace*{-0.26\textwidth}(e)\hspace*{0.26\textwidth}
    \includegraphics[height=0.172\textwidth,valign=t]{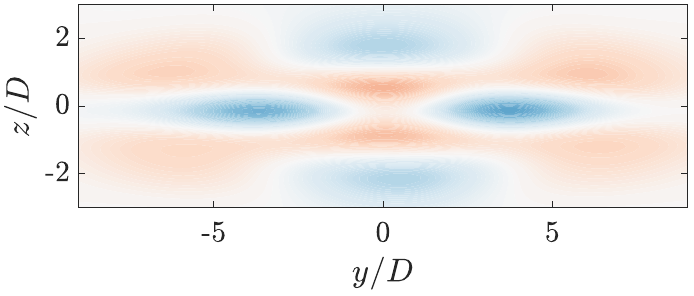} 
    \hspace*{-0.41\textwidth}(f)\hspace*{0.42\textwidth}~\\
    \vspace{2mm}
    \includegraphics[height=0.042\textwidth]{Results/12_Gauss/GaussDiff_colorbar.pdf}
    \caption{The difference between the mean stationary velocity profile $\overbar{U}_s(y,z,t)$ and its Gaussian fit in the $y-z$ plane, normalized by the center velocity deficit value $U_{0}$, at the same time steps and cases as in figure \ref{fig:results-gauss-origin}. }
    \label{fig:results-gauss-moved-diff}
\end{figure}

To evaluate the contribution of the meandering in the distortion of mean velocity profile, we examine the normalized difference between the mean stationary velocity $\overbar{U}_s(y,z,t)$ and its 2D Gaussian fit in figure \ref{fig:results-gauss-moved-diff}, instead of the equivalent difference for the mean velocity profile $\overbar{U}(y,z,t)$ in figure \ref{fig:results-gauss-origin-diff}. 
We note that the mean stationary profile $\overbar{U}_s(y,z,t)$ is averaging the velocity profiles without meandering effect, while the mean velocity profile $\overbar{U}(y,z,t)$ is averaging the velocity profiles with meandering effect. 
They are bridged through the conditional stationary profile in equations \eqref{eqn:model-Us} and \eqref{eqn:model-superposition}. 
In figure \ref{fig:results-gauss-moved-diff}, we see a pattern of the growing distortion away from the Gaussian behavior, for both cases when the flow evolves, similar to figure \ref{fig:results-gauss-origin-diff}. 
The vertical centerline profiles still have a thicker tail than the horizontal centerline profiles. 
However, the general level of deviation is greatly reduced. 
In the Q2D regime, the maximum absolute value of the normalized deviation is 5.6\%, and the areas of large deviation are reduced to 0.22$D^2$ and 0.28$D^2$ in figures \ref{fig:results-gauss-moved-diff} (c,f) respectively. 
The deviation is now qualitatively mainly restrained in the wake core after excluding the meandering effect. 
Therefore, the meandering effect, as the main difference between $U$ and $\overbar{U}_s$, is one major source of the distortion of self-similarity.
We will further quantify the meandering behavior in the following sections and evaluate its impact on the scaling of the statistics, such as the velocity deficit.

\subsection{Meandering behavior} \label{sub:results-pdf}

\begin{figure}
    \centering
    \includegraphics[height=0.14\textwidth]{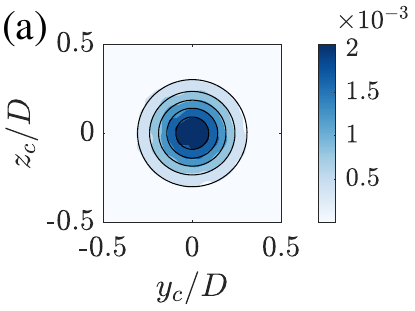}
    \includegraphics[height=0.14\textwidth]{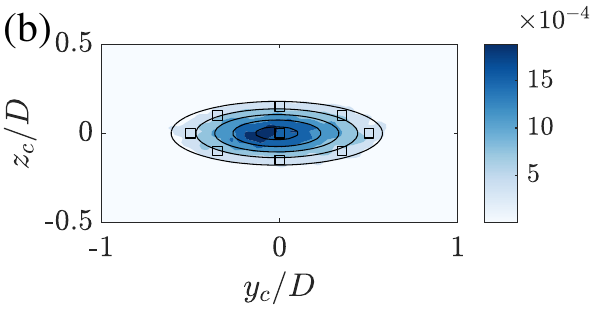}
    \includegraphics[height=0.14\textwidth]{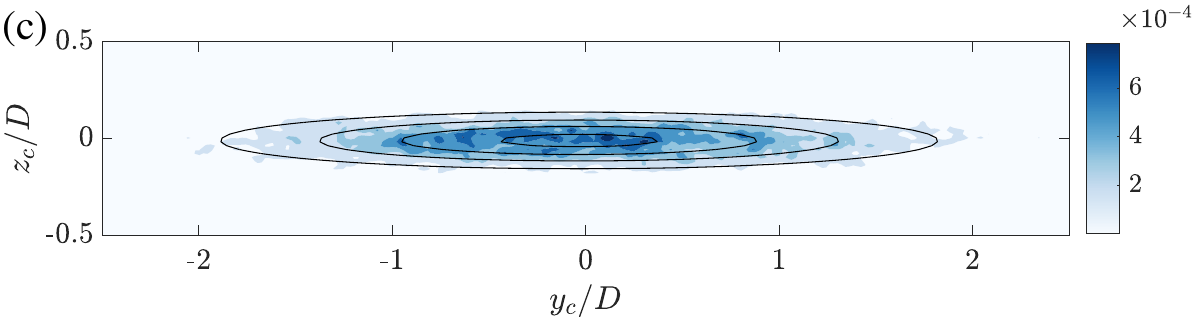}
    \includegraphics[height=0.14\textwidth]{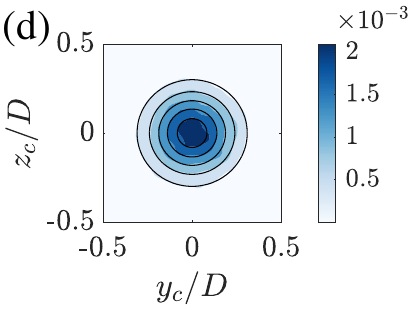}
    \includegraphics[height=0.14\textwidth]{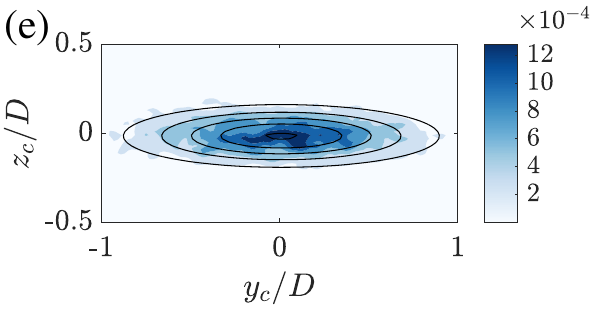}
    \includegraphics[height=0.14\textwidth]{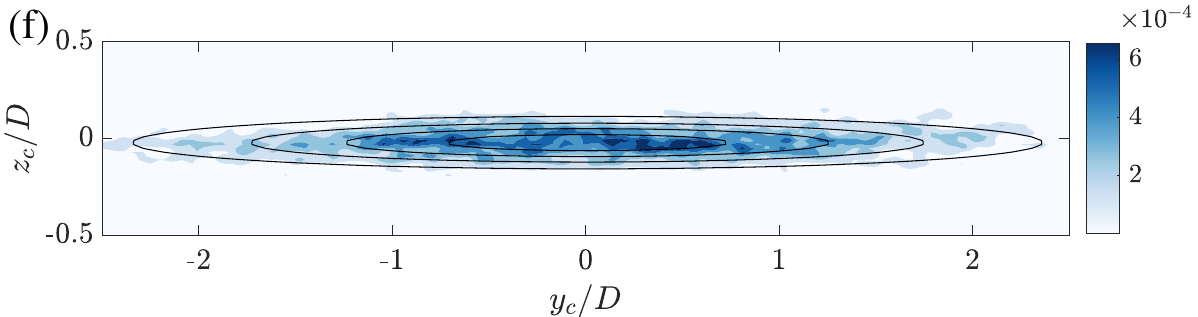}
    \caption{The joint p.d.f. of the instantaneous center location $(y_c, z_c)$ in the $y-z$ plane for (a,b,c) R20F02 and (d,e,f) R20F50. 
    The solid lines are the same contour levels from the 2D Gaussian fit of the joint p.d.f. $p_c(y_c,z_c,t)$. 
    The examined time steps are the same as figure \ref{fig:results-gauss-origin}. 
    In (b), the blue squares are the center locations of the conditional stationary velocity profile $\overbar{U}_s$ in figure \ref{fig:results-cond2d}. }
    \label{fig:results-gauss-ctrpdf}
\end{figure}

To quantify the meandering behavior of the stratified wake, we extract the center locations and the width and height of the instantaneous velocity profiles. 
We focus on the corresponding joint probability density functions of the meandering center location $(y_c,z_c)$ and the wake width and height $(W,H)$, $p_c(y_c,z_c,t)$ and $p_l(W,H,t)$ for all three regimes in both R20F02 and R20F50 cases. 
The former, $p_c(y_c,z_c,t)$, reveals the range of the meandering, and the latter, $p_l(W,H,t)$, measures the possibility of the distortion in the instantaneous velocity profiles during the process of meandering. 
They are relevant to the scaling of the center velocity deficit through $g(\xi_{sc},\eta_{sc})$ and $\hat{p}_c(\xi_{pc},\eta_{pc})$ in equation \eqref{eqn:model-ss-U0superpose} respectively. 

The joint p.d.f. of the center location $p_c(y_c,z_c,t)$ is presented in figure \ref{fig:results-gauss-ctrpdf}.
As the flow develops downstream, $p_c(y_c,z_c,t)$ is expanding fast in the horizontal direction, similar to the wake width $W$. 
This indicates a simultaneous horizontal expansion of the meandering range along with the widening of the velocity profile. 
On the other hand, the range of meandering continuously shrinks in the vertical direction from the NEQ regime to the Q2D regime, although the wake height $H$ is maintained near a constant as shown in figure \ref{fig:results-gauss-origin}. 
The largely shrinking vertical meandering along with the plateau in the wake height qualitatively implies that stratification suppresses meandering strongly compared to its impact on the velocity profile. 
This difference will be quantitatively examined in Sec. \ref{sub:results-sp} in the scaling of statistics. 

As a result, both the width and the height of $p_c(y_c,z_c,t)$ for R20F02 are comparable to R20F50, regardless of the background stratification strength $Fr_B$. 
The case R20F50 has a slightly narrower vertical meandering range while it has a slightly larger horizontal meandering range, as shown in figures \ref{fig:results-gauss-ctrpdf} (c,f). 
We note that the range of both the horizontal and the vertical meandering is much smaller than the width and height of the wake velocity profile. 
According to equations \eqref{eqn:model-ss-Gauss-length}, the superposed width and height are quadratically impacted by the width and height of the meandering range and the mean stationary velocity profile, and the relatively small range of meandering will constrain the impact of the meandering.

\begin{figure}
    \centering
    \includegraphics[height=0.20\textwidth,valign=t]{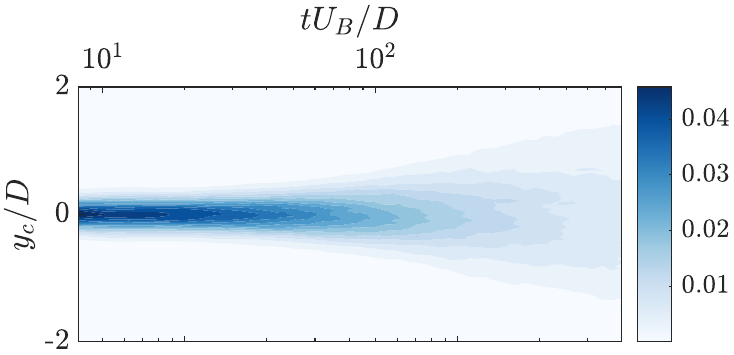}\hspace{3mm}
    \hspace*{-0.43\textwidth}(a)\hspace*{0.42\textwidth}
    \includegraphics[height=0.20\textwidth,valign=t]{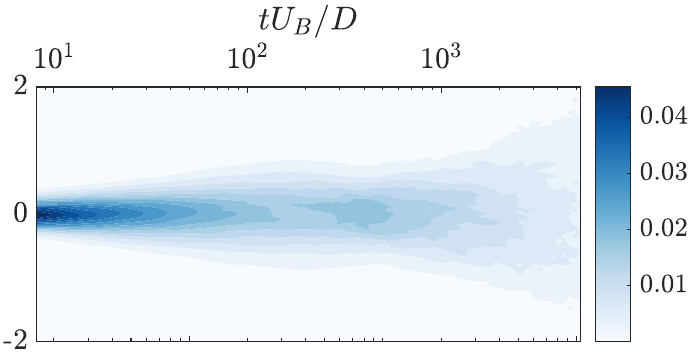}
    \hspace*{-0.42\textwidth}(b)\hspace*{0.45\textwidth}~\\ \vspace{2mm}
    (c)\hspace*{0.38\textwidth}
    (d)\hspace*{0.44\textwidth}~\\
    \vspace{1mm}
    \includegraphics[height=0.175\textwidth,valign=t]{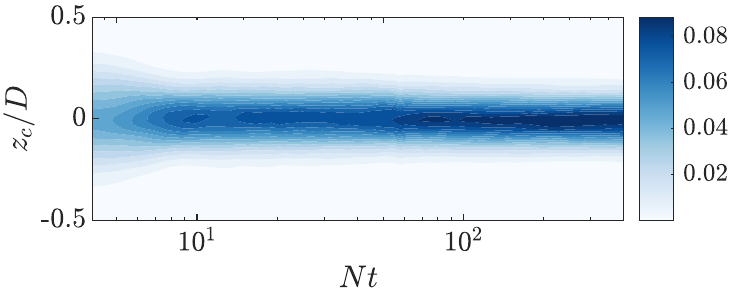}\hspace{1mm}
    \includegraphics[height=0.175\textwidth,valign=t]{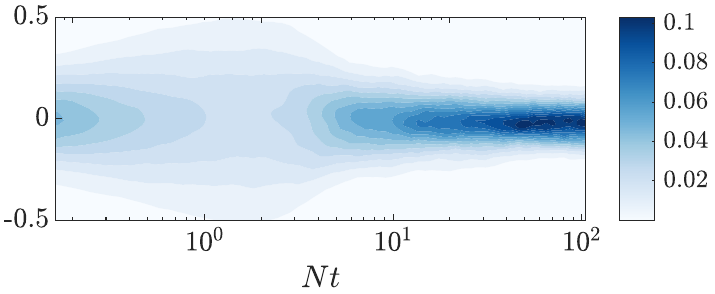} \hspace*{0.04\textwidth}
    \caption{The development of the marginal p.d.f. of the center location in (a,b) the $y$-direction $p_{c,Y}(y_c,t)$ and (c,d) in the $z$-direction $p_{c,Z}(z_c,t)$. 
    The left column (a,c) are for the case R20F02, and the right column (b,d) are for the case R20F50. }
    \label{fig:results-gauss-ctrmarg}
\end{figure}

To visualize the development of the meandering range for $y$- and $z$-directions respectively, the marginal p.d.f. profiles of the wake center locations are shown in figure \ref{fig:results-gauss-ctrmarg} for both cases R20F02 and R20F50. 
The marginal p.d.f. profiles of the horizontal location $p_{c,Y}(y_c,t)$ and the vertical location $p_{c,Z}(z_c,t)$ are examined as a function of the non-dimensional time, $Nt$, where the subscript $Y$ or $Z$ indicates the spatial marginal direction. 
The widths of the profiles $p_{c,Y}(y_c,t)$ and $p_{c,Z}(z_c,t)$ suggest the range of meandering in the horizontal $y$- and the vertical $z$-directions. 
In general, the range of the horizontal meandering keeps growing through the wake in figure \ref{fig:results-gauss-ctrmarg} (a,b), while the range of the vertical meandering collapses in the late wake in figure \ref{fig:results-gauss-ctrmarg} (c,d). 
To be specific, we observe roughly three stages out of the development of the meandering range in the cases. 
For the horizontal meandering, the range is growing at a decaying rate before $Nt\sim 3$. 
The background stratification in R20F02 is so strong that there is almost no such stage. 
When $Nt$ is between $[3, 20]$, the horizontal range of meandering does not change much and reaches a plateau for both cases in figure \ref{fig:results-gauss-ctrmarg} (a,b). 
After $Nt\sim 20$, the meandering in the horizontal direction starts to grow again. 
For the vertical meandering, the range is similarly growing at a decaying rate before $Nt\sim 2$. 
When $Nt$ is between $[2, 8]$, the vertical range of meandering gradually decreases for both cases in figure \ref{fig:results-gauss-ctrmarg} (c,d). 
After $Nt\sim 8$, the meandering in the vertical direction is limited to a similar range for both R20F02 and R20F50. 
Note that these three stages do not necessarily correspond to the NW regime, the NEQ regime and the Q2D regime. 
Though both the horizontal meandering range and the vertical meandering range have three stages when evolving, they do not share the same transition points between stages. 
A similar developing pattern of the meandering range in both directions is also observed in other cases. 
This is another evidence of the difference between the horizontal direction and the vertical direction, and urges us to treat the horizontal direction and the vertical direction differently during the analysis of scaling in Sec. \ref{sub:results-sp}.

\begin{figure}
    \centering
    \includegraphics[height=0.17\textwidth,valign=t]{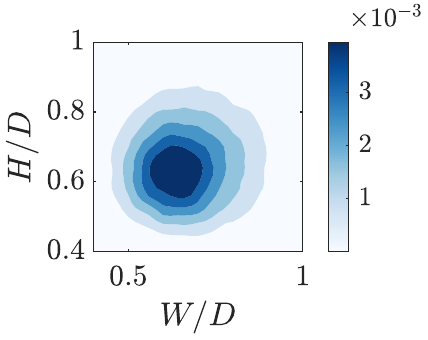}
    \hspace*{-0.22\textwidth}(a)\hspace*{0.19\textwidth}
    \includegraphics[height=0.17\textwidth,valign=t]{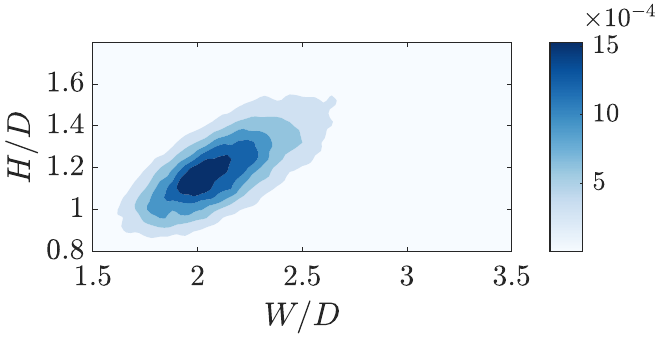}
    \hspace*{-0.33\textwidth}(b)\hspace*{0.3\textwidth}
    \includegraphics[height=0.17\textwidth,valign=t]{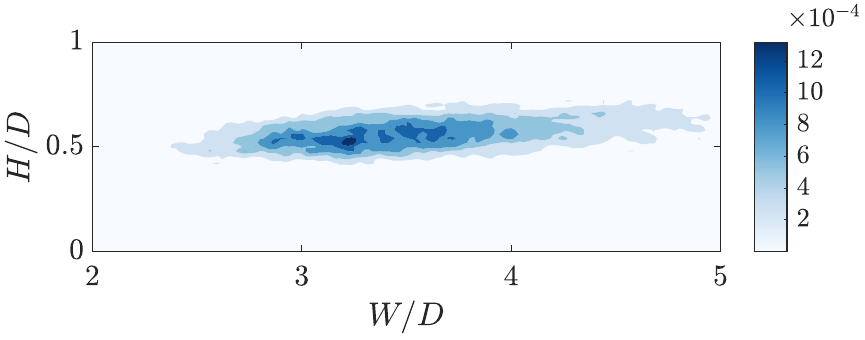}
    \hspace*{-0.44\textwidth}(c)\hspace*{0.41\textwidth}~\\
    \includegraphics[height=0.17\textwidth,valign=t]{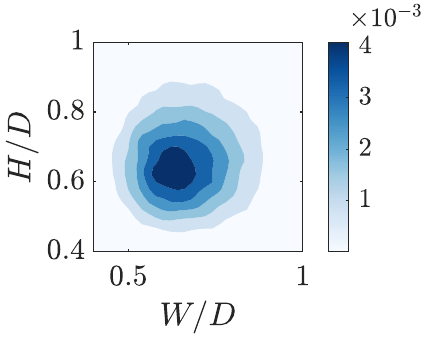}
    \hspace*{-0.22\textwidth}(d)\hspace*{0.19\textwidth}
    \includegraphics[height=0.17\textwidth,valign=t]{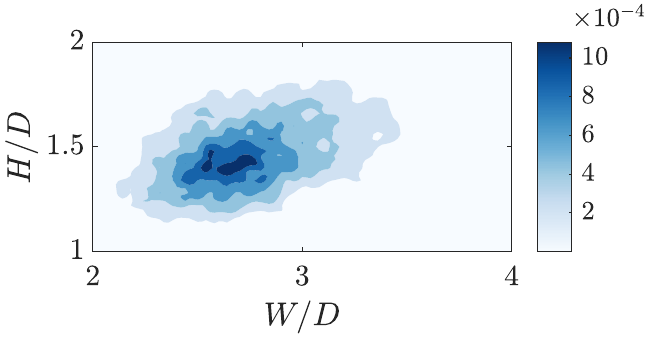}
    \hspace*{-0.33\textwidth}(e)\hspace*{0.3\textwidth}
    \includegraphics[height=0.17\textwidth,valign=t]{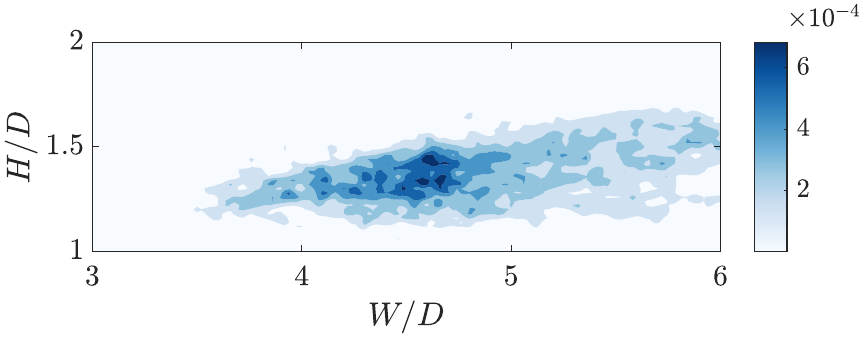}
    \hspace*{-0.44\textwidth}(f)\hspace*{0.41\textwidth}~\\
    \caption{The joint p.d.f. of the instantaneous width and height $(W,H)$ for the same cases and the same time steps as in figure \ref{fig:results-gauss-origin}. }
    \label{fig:results-length-jointpdf}
\end{figure}

Figure \ref{fig:results-length-jointpdf} shows how the joint p.d.f. of the instantaneous width and height $p_l(W,H,t)$ changes over time. 
$p_l(W,H,t)$ measures the possibility of distortion in the instantaneous velocity profiles during the process of meandering. 
The impact of distortion on velocity scaling will be partially reflected through $g(\xi_{sc},\eta_{sc})$ in equation \eqref{eqn:model-ss-U0superpose} under the self-similarity assumption. 
In the NW regime in figure \ref{fig:results-length-jointpdf} (a,d), both the width $W$ and the height $H$ span a wide range and they are comparable. 
They are positively correlated, though the correlation is not strong. 
The length in the horizontal direction, $W$, still performs similarly as the length in the vertical direction, $H$. 
In the NEQ regime in figure \ref{fig:results-length-jointpdf} (b,e), both $W$ and $H$ grow larger, though the width $W$ is much larger than the height $H$. 
Both $W$ and $H$ span larger in distribution, and the positive correlation grows stronger. 
This means that an instantaneous velocity profile that expands in the horizontal direction is more likely to expand in the vertical direction. 
However, $W$ varies in a wider range than $H$, and thus, the horizontal direction gradually differs from the vertical direction. 
In the Q2D regime in figure \ref{fig:results-length-jointpdf} (c,f), the positive correlation is weak. 
The wake height $H$ is close to a constant value, especially compared to the wide span in p.d.f. for the wake width $W$. 
The horizontal direction is now totally different from the vertical direction regarding the range of variation. 
Case R20F50 in figure \ref{fig:results-length-jointpdf} (f) shows a larger variation of $H$ than case R20F02 in figure \ref{fig:results-length-jointpdf} (c) due to the weaker background stratification. 

In general, the meandering has a trend of a fast-expanding horizontal range and a continuously shrinking vertical range, shown by the joint p.d.f. of the instantaneous center location $p_c(y_c,z_c,t)$ in figure \ref{fig:results-gauss-ctrpdf}. 
The expanding range of meandering in the horizontal direction leads to a larger variation in the instantaneous width $W$, while the suppressed range of meandering in the vertical direction leads to a small variation in the instantaneous height $H$ as in figure \ref{fig:results-length-jointpdf}. 
The horizontal direction thus differs from the vertical direction in its meandering behavior and also differs in the transition points between the three stages of development in the meandering range. 
The strength of background stratification, proportional to $Fr_B^{-1}$, does not impact the developing pattern of meandering behavior significantly. 
The meandering keeps developing stronger as the wake flow evolves downstream and brings a stronger distortion in the mean velocity profiles in figure \ref{fig:results-gauss-origin-diff}. 

\begin{figure}
    \centering
    (a)~\includegraphics[width=.32\textwidth,valign=t]{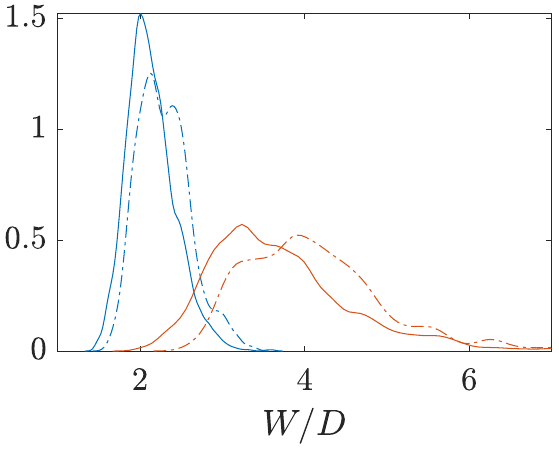}
    ~(b)\includegraphics[width=.32\textwidth,valign=t]{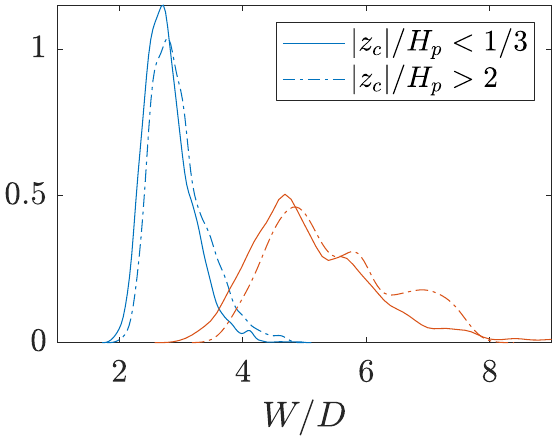}
    \caption{The p.d.f. of the instantaneous width conditioned on a weak vertical meandering, $|z_c|/H_p<1/3$, given by the solid lines, and on a strong vertical meandering, $|z_c|/H_p>2$, given by the dash-dotted lines. 
    (a) Case R20F02. The blue lines are at $Nt=45.6$, and the orange lines are at $Nt=400.3$. 
    (b) Case R20F50. The blue lines are at $Nt=18.6$, and the orange lines are at $Nt=104.6$. }
    \label{fig:results-length-cond}
\end{figure}

To verify the causality between the distortion in the velocity profile and the meandering behavior, we examine the p.d.f. of the width when conditioned on different levels of meandering in figure \ref{fig:results-length-cond}. 
We use solid lines to indicate a weak vertical meandering, i.e., $|z_c|/H_p<1/3$, and dash-dotted lines to indicate a strong vertical meandering, i.e., $|z_c|/H_p>2$. 
The difference between the solid lines and the dash-dotted lines specifies the influence of the meandering behavior. 
The blue lines are in the NEQ regime where the meandering is weaker, while the orange lines are in the Q2D regime where the meandering is stronger. 
Both cases R20F02 and R20F50 are examined, shown in figure \ref{fig:results-length-cond} (a,b) respectively. 
The conditional p.d.f.s have a similar range no matter whether the vertical meandering is weak or strong when the flow has not entered the late wake, as shown by the blue lines. 
Stronger background stratification in R20F02 will lead to another peak with a large width in figure \ref{fig:results-length-cond} (a), though. 
As the flow enters the Q2D regime, the strong vertical meandering significantly increases the wake width, as shown by the orange lines. 
For R20F02, the p.d.f. has a larger peak wake width, while for R20F50, the p.d.f. obtains another peak near the tail. 
The influence of meandering is stronger since the meandering itself is stronger in the Q2D regime. 
The difference between the NEQ regime and the Q2D regime suggests that the strong late wake meandering is a source of the wake profile distortion.

\subsection{Conditional averaging} \label{sub:results-condavg}

\begin{figure}
    \centering
    \includegraphics[width=0.3\textwidth,valign=t]{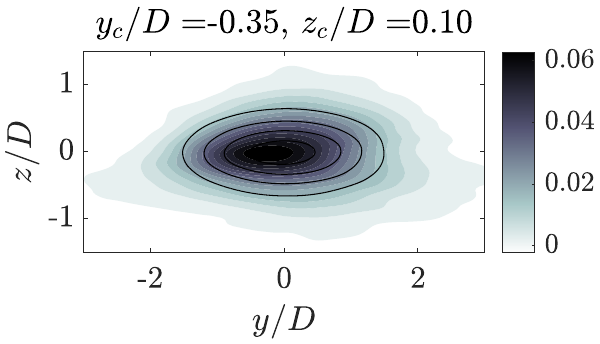}
    \hspace*{-0.3\textwidth}(a)\hspace*{0.26\textwidth}~
    \includegraphics[width=0.3\textwidth,valign=t]{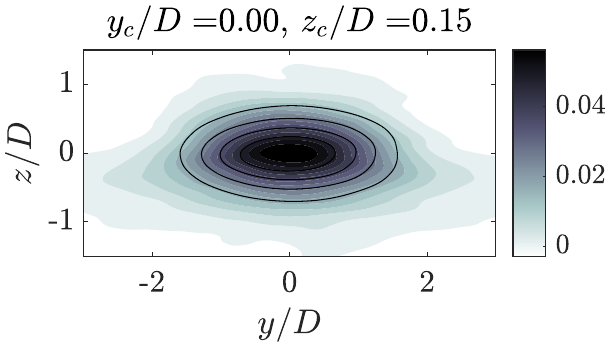}
    \hspace*{-0.3\textwidth}(b)\hspace*{0.26\textwidth}~
    \includegraphics[width=0.3\textwidth,valign=t]{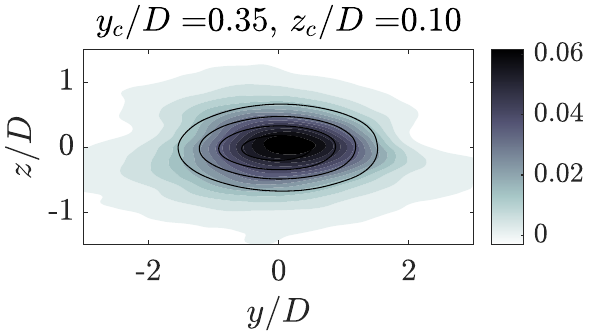}
    \hspace*{-0.3\textwidth}(c)\hspace*{0.26\textwidth}~\\
    \includegraphics[width=0.3\textwidth,valign=t]{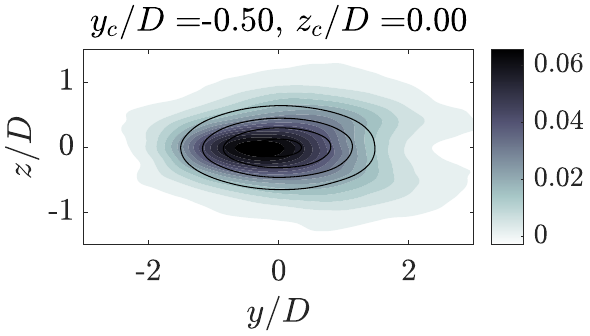}
    \hspace*{-0.3\textwidth}(d)\hspace*{0.26\textwidth}~
    \includegraphics[width=0.3\textwidth,valign=t]{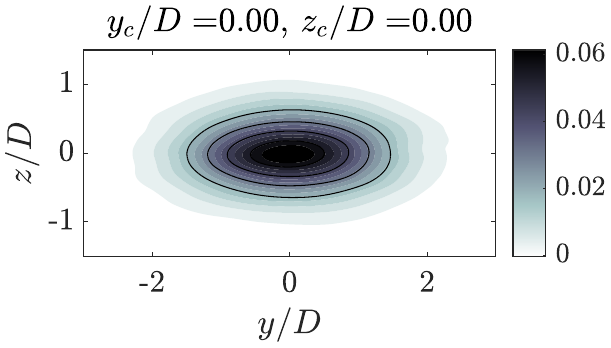}
    \hspace*{-0.3\textwidth}(e)\hspace*{0.26\textwidth}~
    \includegraphics[width=0.3\textwidth,valign=t]{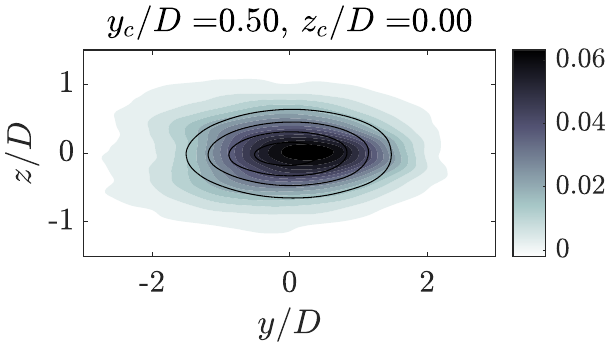}
    \hspace*{-0.3\textwidth}(f)\hspace*{0.26\textwidth}~\\
    \includegraphics[width=0.3\textwidth,valign=t]{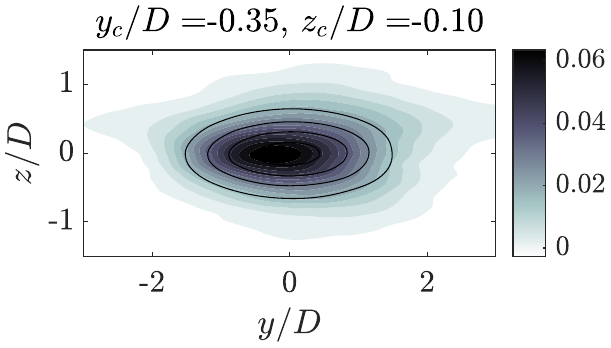}
    \hspace*{-0.3\textwidth}(g)\hspace*{0.26\textwidth}~
    \includegraphics[width=0.3\textwidth,valign=t]{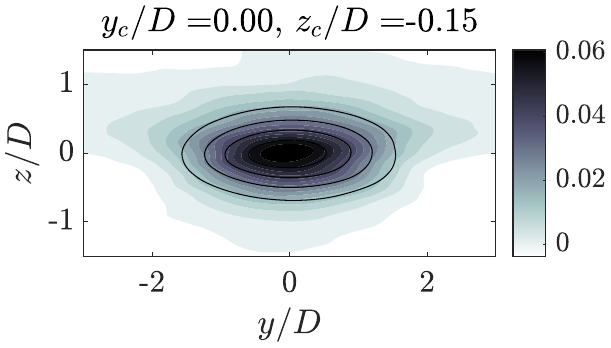}
    \hspace*{-0.3\textwidth}(h)\hspace*{0.26\textwidth}~
    \includegraphics[width=0.3\textwidth,valign=t]{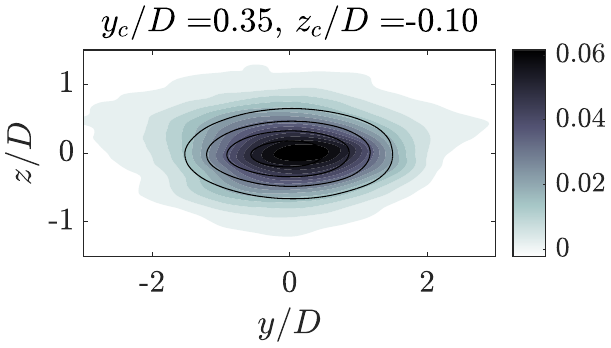}
    \hspace*{-0.3\textwidth}(i)\hspace*{0.26\textwidth}~\\
    \caption{The conditional stationary velocity profile $U_s(y,z,y_c,z_c,t)$ conditionally averaged on the center location $(y_c,z_c)$. 
    The time step is $Nt=45.6$ from R20F02, which falls in the NEQ regime. 
    The examined center locations are also visualized by the blue squares in figure \ref{fig:results-gauss-ctrpdf} (b). 
    Here, the local p.d.f. width is $W_p/D=0.33$, and the local p.d.f. height is $H_p/D=0.10$. 
    The black lines indicate the same contour levels in the mean stationary profile $\overbar{U}_s(y,z,t)$, and only 1 out of every 3 contour lines near the wake core are shown for clarity. }
    \label{fig:results-cond2d}
\end{figure}

Due to the meandering of the wake, the deficit velocity profile $\overbar{U}(y,z,t)$ is a blending of the conditional stationary velocity profile $U_s(y,z,y_c,z_c,t)$ and the surrounding flow. 
The surrounding flow is constant zero since the simulations use the laboratory frame as the reference coordinates, so only $U_s(y,z,y_c,z_c,t)$ and $p_c(y_c,z_c,t)$ show up in equation \eqref{eqn:model-superposition}. 
In Sec. \ref{sub:results-pdf}, we examined the partial impact of distortion on the scaling of the center velocity deficit through $g(\xi_{sc},\eta_{sc})$ and $\hat{p}_c(\xi_{pc},\eta_{pc})$ in equation \eqref{eqn:model-ss-U0superpose}. 
However, the simplified superposition equation \eqref{eqn:model-ss-U0superpose} for center velocity deficit requires self-similarity in the shape of the conditional stationary velocity, i.e., a universal $g(\xi_s,\eta_s)$ in equation \eqref{eqn:model-ss-Usdef}. 
To examine the universality in the velocity profile, here, we turn to examine the conditional stationary velocity profiles $U_s(y,z,y_c,z_c,t)$ obtained by conditional averaging on the center location $(y_c,z_c)$. 
In figure \ref{fig:results-cond2d}, the examined $(y_c,z_c)$ pairs are offset from the mean wake center in different directions, and the offset distance is comparable to the local width $W_p$ and height $H_p$ of the joint p.d.f. $p_c(y_c,z_c,t)$. 

The shape of the conditional stationary velocity profiles $U_s(y,z,y_c,z_c,t)$ measures how distorted an instantaneous velocity profile will be when the wake core meanders to a specific center location $(y_c,z_c)$, which is visualized as the blue squares in figure \ref{fig:results-gauss-ctrpdf} (b). 
In figure \ref{fig:results-cond2d} (e), when $(y_c,z_c)$ is at the mean wake center, $U_s(y,z,y_c,z_c,t)$ is Gaussian like. I
t is almost the same as the mean stationary profile $\overbar{U}_s(y,z,t)$, given by the solid black contour lines. 
However, with $(y_c,z_c)$ away from the mean wake core, $U_s(y,z,y_c,z_c,t)$ is distorted due to the meandering. 

In figures \ref{fig:results-cond2d} (b,h), when $(y_c,z_c)$ is offset from the mean wake center in the vertical direction, i.e., $y_c=0$, $U_s(y,z,y_c,z_c,t)$ is still similar to the mean stationary profile $\overbar{U}_s(y,z,t)$ near the mean wake core. 
When $y_c=0,z_c>0$, the wake meanders to the upper side of the domain, and thus the lower half of the profile $z<0$ is near the mean wake core; 
We see a wider profile in the lower half of the wake in the periphery, and this is due to the stronger turbulence intensity in the mean wake core. 
When $y_c=0,z_c<0$, things are the opposite, and the upper half of the profile $z>0$ is near the mean wake core; 
We see a wider profile in the upper half of the wake in the periphery, instead. 

In figures \ref{fig:results-cond2d} (d,f), when $(y_c,z_c)$ is offset from the mean wake center in the horizontal direction, i.e., $z_c=0$, $U_s(y,z,y_c,z_c,t)$ is distorted both near the wake core and in the periphery. 
Similarly, when $y_c>0,z_c=0$, the conditional stationary profile is vertically thicker in the left half of the profile $y<0$ since it meanders closer to the wake core with more active turbulence; 
When $y_c<0,z_c=0$, the conditional stationary profile is vertically thicker in the right half of the profile $y>0$. 

In figures \ref{fig:results-cond2d} (a,c,g,i), the distortion is the combination of the horizontal offset and the vertical offset. 
We observe that the distortion near the wake core is mainly the asymmetry between the left and the right halves, and the core is close to symmetric between the top and the bottom halves. 
However, the distortion in the periphery is in both the vertical and the horizontal directions, and no symmetry is observed. 
Again, we see that the distortion of the conditional stationary velocity profile is different between the horizontal direction and the vertical direction.

\begin{figure}
    \centering
    \includegraphics[width=0.3\textwidth,valign=t]{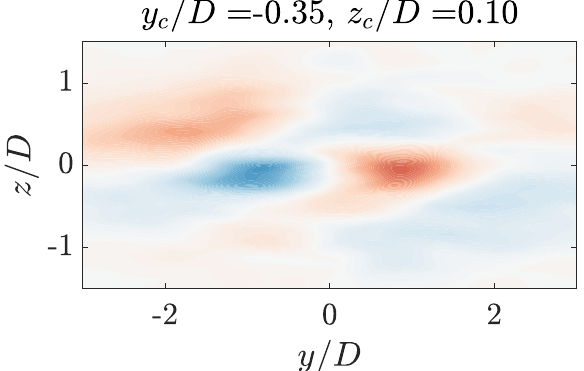}
    \hspace*{-0.3\textwidth}(a)\hspace*{0.26\textwidth}~
    \includegraphics[width=0.3\textwidth,valign=t]{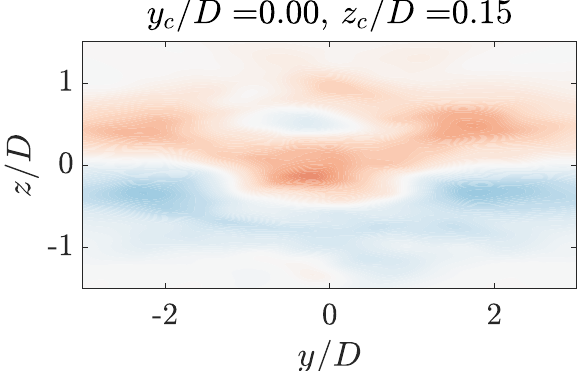}
    \hspace*{-0.3\textwidth}(b)\hspace*{0.26\textwidth}~
    \includegraphics[width=0.3\textwidth,valign=t]{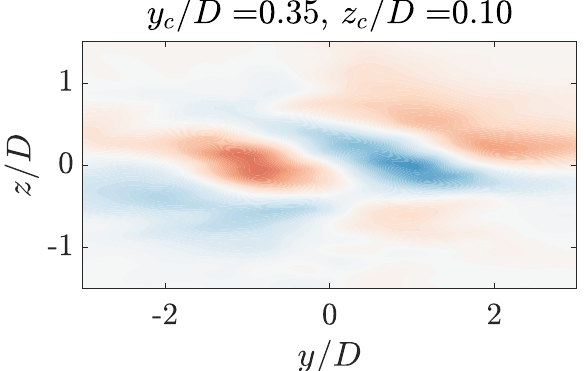}
    \hspace*{-0.3\textwidth}(c)\hspace*{0.26\textwidth}~\\
    \vspace{1mm}
    \includegraphics[width=0.3\textwidth,valign=t]{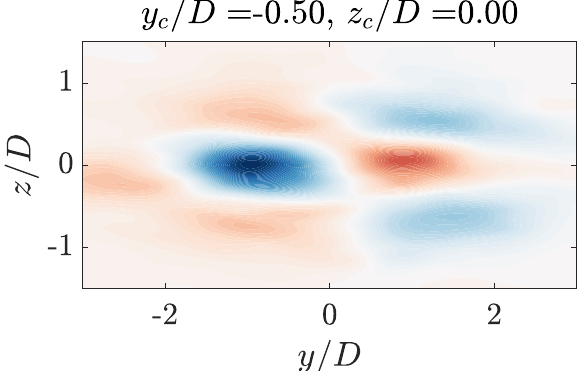}
    \hspace*{-0.3\textwidth}(d)\hspace*{0.26\textwidth}~
    \includegraphics[width=0.3\textwidth,valign=t]{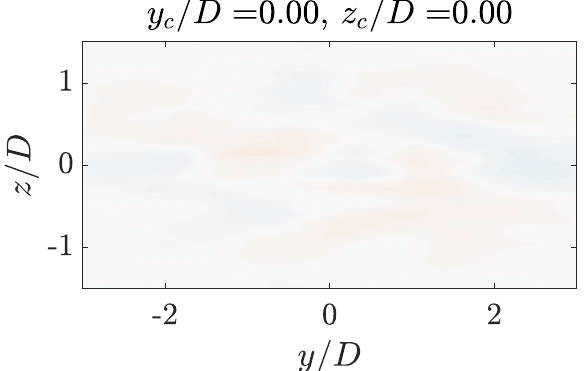}
    \hspace*{-0.3\textwidth}(e)\hspace*{0.26\textwidth}~
    \includegraphics[width=0.3\textwidth,valign=t]{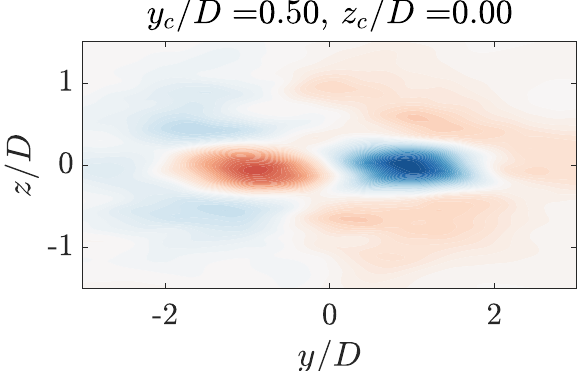}
    \hspace*{-0.3\textwidth}(f)\hspace*{0.26\textwidth}~\\
    \vspace{1mm}
    \includegraphics[width=0.3\textwidth,valign=t]{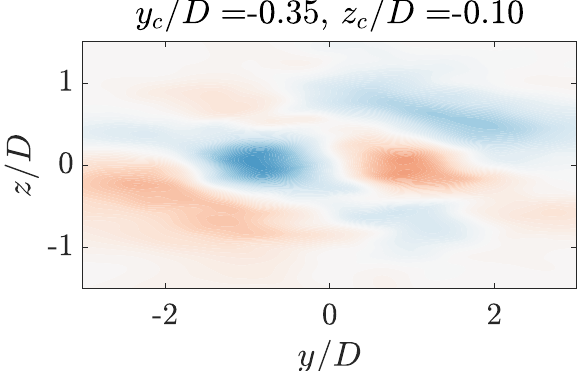}
    \hspace*{-0.3\textwidth}(g)\hspace*{0.26\textwidth}~
    \includegraphics[width=0.3\textwidth,valign=t]{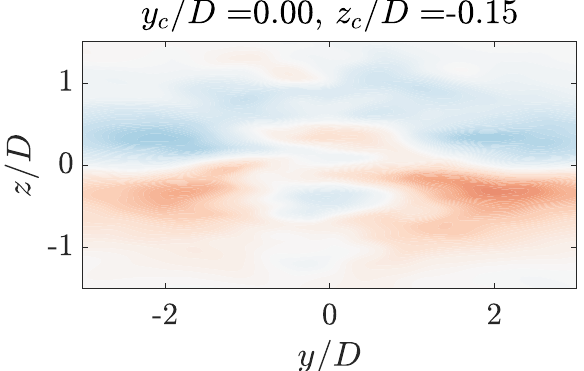}
    \hspace*{-0.3\textwidth}(h)\hspace*{0.26\textwidth}~
    \includegraphics[width=0.3\textwidth,valign=t]{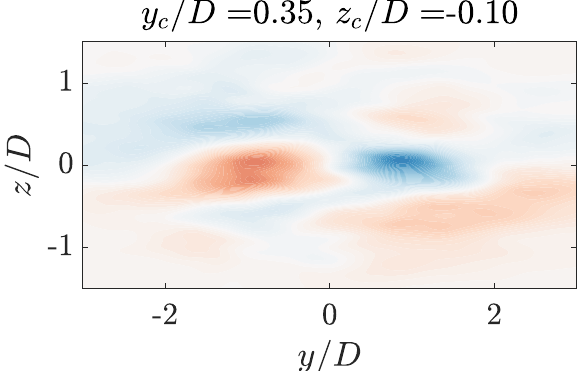} 
    \hspace*{-0.3\textwidth}(i)\hspace*{0.26\textwidth}~\\
    \vspace{3mm}
    \includegraphics[height=0.042\textwidth]{Results/16_conditional/GaussDiff_colorbar.pdf}
    \caption{The difference $\Delta U_s/U_{s0}$ between the conditional stationary velocity $U_s(y,z,y_c,z_c,t)$ and the mean stationary velocity $\overbar{U}_s(y,z,t)$, normalized by $U_{s0}$. 
    The examined center locations and the time step are the same as in figure \ref{fig:results-cond2d}. 
    Here, the local p.d.f. width is $W_p/D=0.33$, and the local p.d.f. height is $H_p/D=0.10$. }
    \label{fig:results-cond2d-diff}
\end{figure}

Next, we will visualize the extra structures in addition to the mean stationary profile that leads to the difference between the horizontal direction and the vertical direction. 
In figure \ref{fig:results-cond2d-diff}, we show the normalized difference $\Delta U_s/U_{s0}$ between the conditional stationary velocity $U_s(y,z,y_c,z_c,t)$ and its mean $\overbar{U}_s(y,z,t)$. 
With the mean $\overbar{U}_s(y,z,t)$ extracted, the difference field $\Delta U_s/U_{s0}$ shows the embedded large-scale structures during meandering. 
We use the local p.d.f. width $W_p/D=0.33$ and height $H_p/D=0.10$ to normalize the meandering distance. 
The blue areas indicate a higher conditional stationary velocity, while the red areas indicate a lower conditional stationary velocity. 

When the meandering is a large horizontal offset, i.e., $|y_c|/W_p=1.5$ as in figure \ref{fig:results-cond2d-diff} (d,f), near the horizontal centerline $z=0$, we see a high-velocity zone in the right half when $y_c>0$ and the left half when $y_c<0$, and a low-velocity zone in the left half when $y_c>0$ and the right half when $y_c<0$. 
Furthermore, away from the horizontal centerline $z\neq 0$, the pair of high-velocity and low-velocity zones switch sides. 
Multiple pairs of high-velocity and low-velocity zones constitute vertically layered vortical structures in $\Delta U_s/U_{s0}$, and the strongest pair is near the center. 
However, when the meandering is a large vertical offset $|z_c|/H_p=1.5$ as in figure \ref{fig:results-cond2d-diff} (b,h), the pairs of high-velocity and low-velocity zones are piled in the vertical direction, and they are not as strong. 
When the center location offset is a combination of $|y_c|/W_p=1.0$ and $|z_c|/H_p=1.0$ as in figure \ref{fig:results-cond2d-diff} (a,c,g,i), $\Delta U_s/U_{s0}$ is mainly vertically layered vortical structures as the pure horizontal offset, with a small tilting of the pairs of high-velocity and low-velocity zones. 
According to equation \eqref{eqn:model-superposition}, the pairs of high-velocity and low-velocity zones give rise to a bit wider profile near the centerlines in both horizontal and vertical directions and a thicker tail in the vertical direction, as shown in figure \ref{fig:results-gauss-origin-marg}. 
Overall, the distortion in the conditional stationary velocity leads to the deviation of the mean stationary velocity $\overbar{U}_s(y,z,t)$ from the Gaussian profile. 
It is also related to the scaling of the centerline deficit velocity through the disruption of the universal $g(\xi_s,\eta_s)$ in equation \eqref{eqn:model-ss-Usdef}.

\subsection{Momentum flux} \label{sub:results-momcons}

\begin{figure}
    \centering
    \includegraphics[width=0.41\textwidth,valign=t]{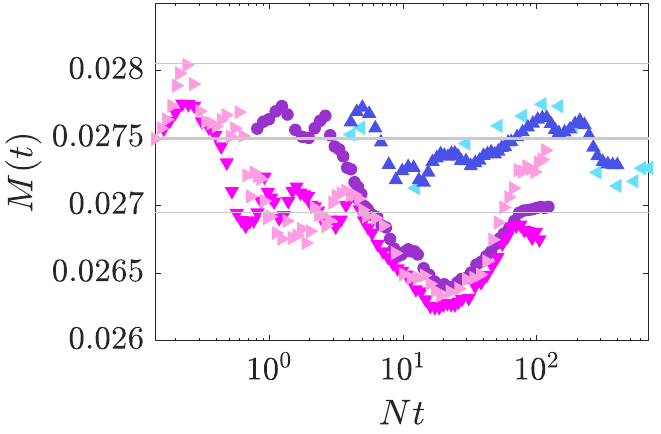}
    \hspace*{-0.42\textwidth}(a)\hspace*{0.38\textwidth}~~~
    \includegraphics[width=0.41\textwidth,valign=t]{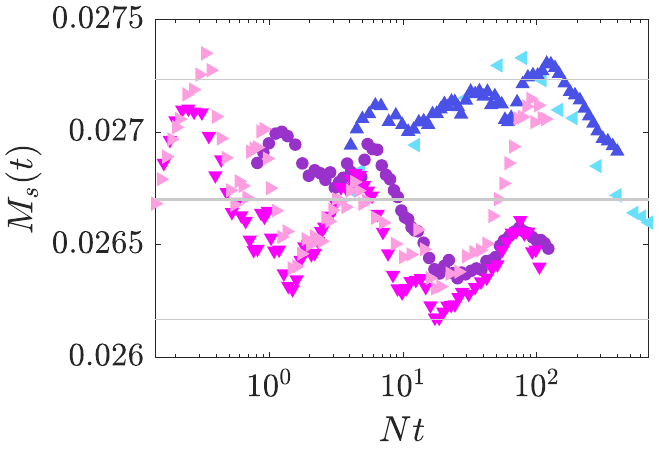}
    \hspace*{-0.42\textwidth}(b)\hspace*{0.38\textwidth}
    \caption{Verification of the conservative equations (a) for $U$, $M(t)={\rm const}$ and (b) for $\overbar{U}_{s}$, $M_s(t)={\rm const}$. 
    The initialization leads to the initial constant value of $M(t)$ to be $M_0=0.275$, indicated by the thick solid line in (a). 
    The thin solid lines are where $102\%, 98\%$ of $M_0$ lie. 
    Similarly, in (b), we indicate the initial value of $M_s(t)$, i.e., $M_{s0}=0.0267$ by the thick solid line, and $102\%, 98\%$ of $M_{s0}$ by the thin solid lines. 
    The color and type of the symbols indicates different cases as defined in table \ref{tab:sim-details}. }
    \label{fig:results-moment-conserv}
\end{figure}

Apart from the distortion in the velocity profile, the meandering behavior also differentiates the statistics of the horizontal direction from those of the vertical direction as shown in the previous sections. 
Not only do the width and the height develop in different patterns, but the horizontal meandering range and the vertical meandering range are also different. 
Nonetheless, the three stages in the developing pattern transition at different points between horizontal meandering and vertical meandering. 
This will change the scaling of the width and the height, and thus will further impact the scaling of the centerline velocity deficit according to the simplified momentum flux equations \eqref{eqn:model-ss-momcons2} and \eqref{eqn:model-ss-momstation}. 

We first verify the simplified momentum flux equations \eqref{eqn:model-ss-momcons2} and \eqref{eqn:model-ss-momstation} in figure \ref{fig:results-moment-conserv}. 
All the cases are plotted, and they are color-coded as in table \ref{tab:sim-details}. 
The nondimensional momentum flux $M(t) = U_{0} WH/U_B D^2$ stands within the range of $95\%\sim 102\% M_0$ for the mean velocity $\overbar{U}(y,z,t)$ and $M_s(t) = U_{s0} W_sH_s/U_B D^2$ mostly within the range of $98\%\sim 102\% M_{s0}$ for the mean stationary velocity $\overbar{U}_s(y,z,t)$. 
Here, $M_0=0.0275$ is the initial nondimensional momentum flux of $\overbar{U}$ for all cases. 
For $\overbar{U}_s$, the initial nondimensional momentum flux $M_{s0}$ is around $0.0267$. 
$M_0$ and $M_{s0}$ are determined during the initialization of the simulations. 

The simplified conservation equations are based on the assumption $U_{0}\ll U_B$ and the assumption of self-similarity in $\overbar{U}$ and $\overbar{U}_s$. 
In the near wake regime, all the cases see a relatively large variation in $M(t)$ and $M_s(t)$ since $U_{0}\sim 0.1U_B$. 
Later, $U_{0}$ gradually decays to a relatively small value. 
The variation of $M(t)$ and $M_s(t)$ is also smaller. 
However, as the deviation from the Gaussian profile grows, $M(t)$ and $M_s(t)$ go through a large variation again. 
Generally, the varying range of $M_s(t)$ is much smaller than the varying range of $M(t)$. 
Hence, the meandering behavior is a factor of the momentum flux deviating from a constant, which is possibly due to its distorting self-similar profiles. 

In the range of $Re_B$ we examine, $Re_B$ does not show its impact on $M(t)$ or $M_s(t)$ till the very late wake, which is the difference we see between R20F50 and R50F50 near $Nt\sim\mathcal{O}(10^2)$ in both figures. 
This is consistent with the fact that the viscous force gradually becomes comparable to the inertial force and impacts the turbulent motions in the late wake \cite{billant2001self}. 
The strength of background stratification is indicated by $Fr_B$, and it has a stronger impact than $Re_B$. 
In Sec. \ref{sub:results-2dss}, a small $Fr_B$ case, e.g., R20F02, has a stronger background stratification, and thus a more suppressed meandering behavior in the vertical direction. 
It has a smaller discrepancy from the 2D Gaussian profile in the late wake. 
Therefore, a smaller variation in $M(t)$ and $M_s(t)$ is also observed for low $Fr_B$ cases. 

In general, the nondimensional momentum flux, $M(t)$ and $M_s(t)$, can be seen as a constant through the wake development. 
Using the momentum flux from the stationary velocity $M_s(t)$ with low $Fr_B$ and high $Re_B$ maximizes the accuracy of that constant. 
The momentum flux equations bridge the scaling of the velocity deficit $U_{0}$ and the scaling of the lengths $W,H$ \cite{meunier2006self}. 
Note that the momentum flux equations \eqref{eqn:model-ss-momcons2} and \eqref{eqn:model-ss-momstation} are obtained based on only two assumptions, and they may also stand in other wakes as long as we have $U_{0}\ll U_B$ and a roughly universal self-similarity.

\subsection{Superposition of the meandering effect} \label{sub:results-sp}

\begin{figure}
    \centering
    \includegraphics[width=0.41\textwidth,valign=t]{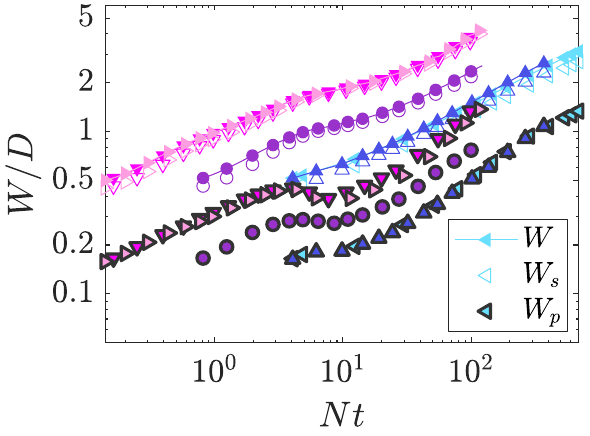}
    \hspace*{-0.42\textwidth}(a)\hspace*{0.38\textwidth}~~~
    \includegraphics[width=0.41\textwidth,valign=t]{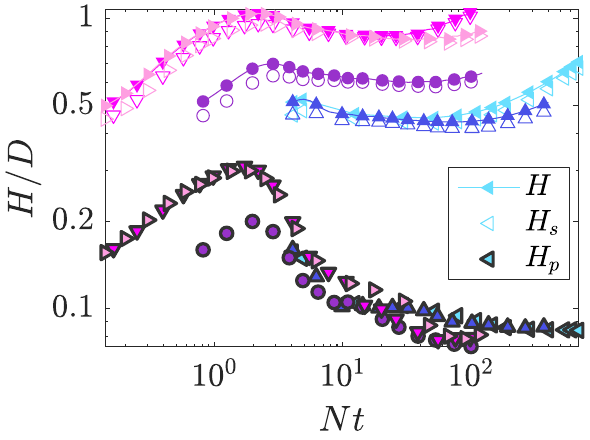}
    \hspace*{-0.42\textwidth}(b)\hspace*{0.38\textwidth}
    \caption{The development of the width and the height. 
    The solid symbols on a solid line are the width and the height of the mean velocity profiles $\overbar{U}$, the hollow symbols are for the mean stationary velocity $\overbar{U}_s$, and the solid symbols with a black outline are for center location p.d.f. $p_c$. 
    The color and type of the symbols indicates different cases as defined in table \ref{tab:sim-details}. 
    Note that all the cases except R10F02 show the symbols every 4 points for clarity. }
    \label{fig:results-superpose-length}
\end{figure}

\begin{figure}
    \centering
    \includegraphics[width=0.40\textwidth,valign=t]{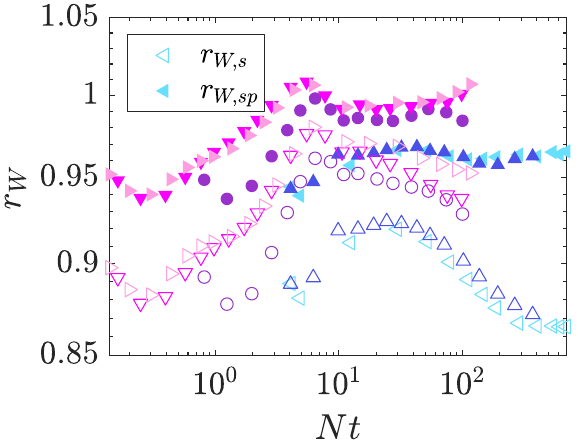}
    \hspace*{-0.42\textwidth}(a)\hspace*{0.38\textwidth}\hspace{5mm}
    \includegraphics[width=0.40\textwidth,valign=t]{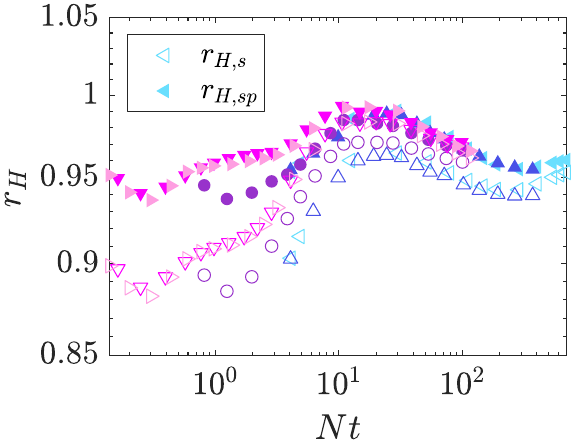}
    \hspace*{-0.42\textwidth}(b)\hspace*{0.38\textwidth}
    \caption{The ratio of (a) the widths and (b) the heights between different length scales. 
    The hollow symbols are for $r_{W/H,s}$ and the solid symbols are for $r_{W/H,sp}$, as defined in equations \eqref{eqn:results-ratio-L}. 
    The color and type of the symbols indicates different cases as defined in table \ref{tab:sim-details}. 
    Note that all the cases except R10F02 show the symbols every 4 points for clarity. }
    \label{fig:results-sp-length-ratio}
\end{figure}

Now, we examine the meandering effect in the statistics scaling. 
The meandering range, quantified by $p_c(y_c,z_c,t)$, can be superposed on the conditional stationary velocity $U_s(y,z,y_c,z_c,t)$ as in equation \eqref{eqn:model-superposition}. 

According to equation \eqref{eqn:model-ss-Gauss-length}, the width and height of the mean profile $W, H$ can be the superposition of the width and height of the mean stationary profile $W_s, H_s$ and the width and height of the center location p.d.f. $W_p, H_p$. 
In figure \ref{fig:results-superpose-length}, the development of multiple widths and heights is shown. 
The widths of the stationary velocity $W_s$ and the meandering range $W_p$ are constantly growing, while each has three stages with different growth rates. 
A clear plateau exists in the meandering width $W_p$ for all cases, indicating a sudden period of suppression in the meandering behavior. 
Compared to the horizontal direction, the vertical lengths, i.e., the heights, show very different patterns in developing. 
First, the height of the stationary velocity $H_s$ grows at the beginning, reaches a plateau, and then grows again at a late stage. 
The first transition of the height $H_s$ scaling is around $Nt\sim 2$, and is earlier compared to $Nt\sim 3$ for the width. 
The following regime of constant height lasts longer till $Nt\sim 50$, which is $Nt\sim 20$ in the width, instead. 
Unlike $H_s$, the height of the meandering range $H_p$ grows at the beginning till $Nt\sim 2$, then decays fast for a short period till $Nt\sim 8$ and decays slowly for a long period until it gradually approaches a constant in the late wake. 
The diverse scaling behavior suggests multiple critical points in the lengths $W_s, W_p, H_s, H_p$, because the impact of meandering is gradually changing through the wake. 
Therefore, the transition points of the four lengths are not necessarily the same as each other. 
The same phenomenon is observed in Sec. \ref{sub:results-pdf} for meandering alone. 

Despite the difference in the transition points among the four lengths, the five cases share the three-stage behavior in all four lengths, and the growth rates in the same regime are close to each other \cite{meunier2006self}. 
In the range of $Re_B$ we examined, $Re_B$ has a minor impact on the widths $W_s$ and $W_p$ through the wake. 
Meanwhile, the heights $H_s$ and $H_p$ do not change with $Re_B$ at the beginning, while low $Re_B$ increases the height $H_s$ greatly in the late wake, which further verifies the importance of the viscous force in the late wake scaling. 
On the other hand, the background Froude number $Fr_B$ changes the length magnitude greatly. 
With stronger background stratification, i.e., low $Fr_B$, the lengths will get earlier affected by the buoyancy, and thus, they usually are a lot smaller at the same $Nt=Fr_B^{-1}\cdot \overbar{x}/D$. 

After analyzing the development stages of the lengths, we now estimate the impact of meandering. 
According to equation \eqref{eqn:model-ss-Gauss-length}, we obtained the superposed width and height $W_{sp}, H_{sp}$. 
To evaluate the accuracy of the superposition model, in figure \ref{fig:results-sp-length-ratio}, we show the length ratios 
\begin{equation}
    r_{W,s} = W_s/W,~ r_{W,sp} = W_{sp}/W,~~~
    r_{H,s} = H_s/H,~ r_{H,sp} = H_{sp}/H. 
    \label{eqn:results-ratio-L}
\end{equation}
If the values of the length ratios are closer to 1, it indicates that the lengths are comparable. 
In general, the width and height of the mean stationary profile $W_s, H_s$ are comparable to those of the mean profile $W, H$, shown by the values of $r_{W,s}$ and $r_{H,s}$. 
However, the meandering ranges, given by the width and height of the center location p.d.f. $W_p, H_p$, are much smaller than the width and the height. 
Since the range of the meandering is around $1/3$ of the mean wake size, the change of the width and height due to the superposition of the meandering will be around 5\%. 
The superposed width and height fall close to the mean wake width and height. 
The rest of the discrepancy, shown by the values of $r_{W,sp}$ and $r_{H,sp}$, shall be attributed to the distortion of the wake profiles, as discussed in the previous sections. 
We observe that the meandering has a consistent impact on the wake width, but a reducing impact on the wake height due to the reducing vertical meandering. 

\begin{figure}
    \centering
    \includegraphics[width=0.43\textwidth,valign=t]{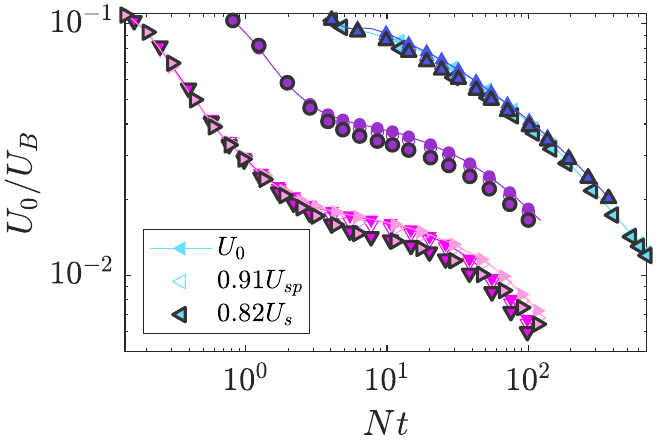}
    \hspace*{-0.44\textwidth}(a)\hspace*{0.40\textwidth}\hspace{5mm}
    \includegraphics[width=0.43\textwidth,valign=t]{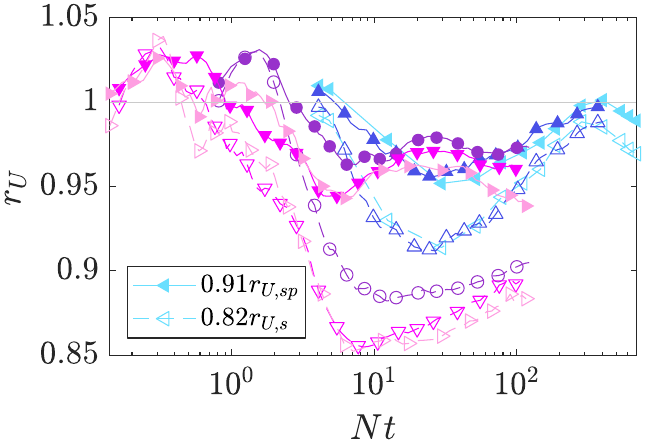}
    \hspace*{-0.45\textwidth}(b)\hspace*{0.41\textwidth}
    \caption{(a) The development of the velocity deficit. 
    The solid symbols on a solid line are the mean velocity deficit $U_{0}$. 
    The hollow symbols are proportional to the superposed velocity $U_{sp0}\equiv M_0/W_{sp}H_{sp}$, obtained from the momentum flux equation, and $0.91$ is the initial value of $U_0/U_{sp0}$. 
    The solid symbols with a black outline are proportional to the mean stationary velocity $U_{s0}$, where $0.82$ is the initial value of $U_0/U_{s0}$. 
    (b) The ratio between the velocity deficits, $r_{U,s}$ and $r_{U,sp}$, as defined in equation \eqref{eqn:results-ratio-U}. 
    The color and type of the symbols indicates different cases as defined in table \ref{tab:sim-details}. 
    Note that all the cases except R10F02 show the symbols every 4 points for clarity. }
    \label{fig:results-moment-Uratio}
\end{figure}

On the other hand, we also want to learn about the impact of the meandering on the center velocity deficit $U_0$. 
Two ways can be used to identify such impact. 

The first way is to use the simplified superposition equation \eqref{eqn:model-ss-Uint}. 
It will lead us to the analytical expression of the self-similar velocity profile. 
We measured the range of meandering in Sec. \ref{sub:results-pdf}. 
The conditional velocity profile $U_s(y,z,y_c,z_c,t)$ is also found to be distorted in Sec. \ref{sub:results-condavg}, due to the large flow structures during meandering. 
They are both related to the meandering behavior, and both contribute to the scaling of the velocity deficit. 
Their contributions are condensed in the coefficient of equation \eqref{eqn:model-ss-U0superpose-const}.
Thus, the mean velocity deficit $U_{0}(t)$ can be simplified to be proportional to the mean stationary deficit $U_{s0}(t)$. 

The second way is to use the momentum flux equation \eqref{eqn:model-ss-momcons2}. 
Here, the width and the height are obtained through the superposition of the meandering in equation \eqref{eqn:model-ss-Gauss-length}. 
This model separates the contribution of the meandering in the horizontal and the vertical directions, but it also involves more assumptions. 

In figure \ref{fig:results-moment-Uratio}, we visualize the two ways to incorporate the meandering effect in the velocity deficit. 
We compare them with the actual center velocity deficit, and we also use the velocity ratios to measure the accuracy of the two models, 
\begin{equation}
\begin{split}
    r_{U,s} & = U_{s0}/U_{0}, \\
    r_{U,sp} & = U_{sp0}/U_{0}, {\rm ~~~ where~} U_{sp0} = M_0/W_{sp}H_{sp}. 
\end{split}
\label{eqn:results-ratio-U}
\end{equation}

In figure \ref{fig:results-moment-Uratio} (a), the first way to use a constant coefficient is given by the filled symbols with a black outline, which does not collapse well with actual $U_0$, given by the filled symbols on a solid line. 
This suggests that the meandering effect does not necessarily bring a constant shift, though the meandering ranges $W_p, H_p$ share a generally similar trend as the mean stationary width and height $W_s, H_s$. 
Such a naive assumption of a constant shift will lead to up to 15\% error as shown by the dashed line symbols in figure \ref{fig:results-moment-Uratio} (b), especially in the late wake regime due to the accumulation of the error. 

However, by using the momentum flux equation, the meandering effect on the velocity deficit is considered a direct outcome of expanding width and height due to the meandering. 
The impact of meandering is separated into the horizontal direction and the vertical direction through $W_{sp}$ and $H_{sp}$. 
The normalized superposed velocity $U_{sp0}(t)$ collapses well with the actual $U_0$ in figure \ref{fig:results-moment-Uratio} (a), and has less than 5\% discrepancy as in figure \ref{fig:results-moment-Uratio} (b). 

In fact, the proportional relationship \eqref{eqn:model-ss-U0superpose-const} can turn to the same form of $U_{sp0}(t)$ using Gaussian profiles, and we leave out the derivation for brevity. 
Therefore, assuming constant in the proportionality \eqref{eqn:model-ss-U0superpose-const} is a further simplified version of $U_{sp0}(t)$, and is expected to have lower accuracy in understanding meandering. 

The mean velocity deficit $U_{0}(t)$ in all cases shows a universal three-stage behavior. 
In the near wake regime, the velocity decays fast, and gradually transitions to a lower decay rate as the buoyancy effect builds up its influence, while in the vertically collapsed late wake regime, the velocity decays fast again. 
Though the decay rates share universality among cases, the transition is not a sudden one since the change of the decay rates is gradual. 
From the view of the momentum conservation, the width and the height are experiencing the transitions at different $Nt$ since the flow develops anisotropy due to the meandering. 
This is one reason for the intricacy of the velocity scaling. 
Though $Fr_B$ only affects the intercept of a power law scaling, $Re_B$ brings a subtle change of the decay rate in the late wake. 
Therefore, it is reasonable to see that more regimes are discovered apart from the traditional three regimes, i.e., the NW, the NEQ and the Q2D regimes, and that the decay rates differ between cases \cite{brucker2010comparative,bonnier2002experimental}. 
A more delicate velocity scaling law with more than three stages is possible. 
Due to the robustness of the momentum flux equation, the consideration of possible stages in width and height will assist in determining the  scaling of the center velocity deficit. 
The utilization of width and height also separates the horizontal direction from the vertical direction, and directly provides physical insights into the turbulent motion evolution.

\section{Conclusion}
\label{sec:conclusion}

Large horizontal structures accompanied by the suppression of the vertical motions are widely observed in the stratified environment. 
As a canonical case, a stratified wake of a sphere has the large horizontal structures emerging in the late wake, and shows strong meandering of the instantaneous velocity profiles. 
However, the growing meandering effects on the wake statistics are seldom evaluated. 
In this work, we study the meandering behavior by examining $\mathcal{O}(100)$ DNSs, which cover $Re_B$ in $[10\,000, 50\,000]$ and $Fr_B$ in $[2,50]$. 
We decompose the instantaneous velocity into the stationary velocity and the center location meandering. 

The horizontal meandering range constantly grows as the stationary profile width does, while the vertical meandering range grows at the beginning and reduces to a constant in the late wake, which is in contrast with the slowly growing stationary profile height. 
In general, the range of meandering is much smaller than the wake profile size. 
Therefore, the existence of meandering leads to around 5\% change in the wake width and height according to the superposition equations, and for the wake height, the meandering effect becomes more insignificant due to the suppression in vertical meandering. 

During the process of meandering, the velocity profile gradually deviates from the self-similar Gaussian profile. 
The conditional velocity profiles will be distorted accordingly based on the meandering location, and thus the mean wake profile grows larger. 
A horizontal center location offset will lead to a large local horizontal structure, while a vertical offset will not. 
Also, the suppression of the vertical meandering is further related to the suppression of the variation in the instantaneous wake height, while the instantaneous wake width still has a large variation range. 
Such differences between the horizontal direction and the vertical direction result in the distortion of the velocity profile, and thus the mean profile deviates from the self-similarity assumption. 
Isolating the stationary velocity from the meandering behavior achieves a better self-similarity. 

According to our theoretical analysis, based on the self-similarity assumption, the nondimensional momentum flux is nearly constant through the wake. 
We verified the constant flux behavior for all the cases, and the momentum flux of the stationary profile without meandering is closer to a constant as expected. 

In the meantime, the momentum flux equation bridges the scaling of the velocity deficit and the scaling of the wake width and height. 
The results suggest a varying meandering effect on the scaling of the velocity deficit, even when the meandering range is comparably small and does not alter the superposed width and height significantly. 
Therefore, a simple proportionality to the stationary velocity deficit is considered an oversimplification of the meandering effect and brings a large discrepancy between the velocity superposed with the meandering and the mean velocity deficit. 

To understand this discrepancy, we examine the width and height, as inspired by the momentum flux equation. 
We find the three-stage behavior in both the width and the height, but the transition points between stages are fairly different among the four lengths $W_s, H_s, W_p, H_p$. 
This brings intricacy to the velocity scaling. 
The scaling of the superposed velocity obtained from the superposed width and height shows a much smaller discrepancy from the mean velocity deficit. 
Therefore, we conjecture that the scaling of the velocity deficit can be better understood by using the scaling of the width and the height. 
The width and the height do not share the critical transition points and thus the velocity scaling will be more complicated than three stages. 

Our results show that $Re_B$ does not come into effect till the late wake, where the viscous force starts to play a role against the buoyancy. 
The background stratification $Fr_B$ directly determines the entering of the buoyancy and thus changes the intercept of the scaling greatly. 
Small $Fr_B$ has a stronger suppression on the vertical motion, especially the vertical meandering range, and also the variation range of the wake height. 
Further efforts are required to find the quantified dependence on the $Re$ number and the $Fr$ number.

\section*{Acknowledgement}
This work is supported by ONR grant no. N000142012315, and the computation is conducted on Penn State Roar system. 
The authors would like to thank Dr. Xiang I.A. Yang for the fruitful discussion on the results, and Dr. Mogeng Li for the technical support for the visualization and the discussion on the results.

\bibliographystyle{ieeetr}
\bibliography{superposition.bib}

\end{document}